\documentstyle[12pt]{article}
\begin{document}
\newcommand {\ee}{\end{equation}}
\newcommand {\bea}{\begin{eqnarray}}
\newcommand {\eea}{\end{eqnarray}}
\newcommand {\nn}{\nonumber \\}
\newcommand {\Tr}{{\rm Tr\,}}
\newcommand {\tr}{{\rm tr\,}}
\newcommand {\e}{{\rm e}}
\newcommand {\etal}{{\it et al.}}
\newcommand {\m}{\mu}
\newcommand {\n}{\nu}
\newcommand {\pl}{\partial}
\newcommand {\p} {\phi}
\newcommand {\vp}{\varphi}
\newcommand {\vpc}{\varphi_c}
\newcommand {\al}{\alpha}
\newcommand {\be}{\beta}
\newcommand {\ga}{\gamma}
\newcommand {\Ga}{\Gamma}
\newcommand {\x}{\xi}
\newcommand {\ka}{\kappa}
\newcommand {\la}{\lambda}
\newcommand {\La}{\Lambda}
\newcommand {\si}{\sigma}
\newcommand {\Si}{\Sigma}
\newcommand {\th}{\theta}
\newcommand {\Th}{\Theta}
\newcommand {\om}{\omega}
\newcommand {\Om}{\Omega}
\newcommand {\ep}{\epsilon}
\newcommand {\vep}{\varepsilon}
\newcommand {\na}{\nabla}
\newcommand {\del}  {\delta}
\newcommand {\Del}  {\Delta}
\newcommand {\mn}{{\mu\nu}}
\newcommand {\ls}   {{\lambda\sigma}}
\newcommand {\ab}   {{\alpha\beta}}
\newcommand {\half}{ {\frac{1}{2}} }
\newcommand {\third}{ {\frac{1}{3}} }
\newcommand {\fourth} {\frac{1}{4} }
\newcommand {\sixth} {\frac{1}{6} }
\newcommand {\sqg} {\sqrt{g}}
\newcommand {\fg}  {\sqrt[4]{g}}
\newcommand {\invfg}  {\frac{1}{\sqrt[4]{g}}}
\newcommand {\sqZ} {\sqrt{Z}}
\newcommand {\gbar}{\bar{g}}
\newcommand {\sqk} {\sqrt{\kappa}}
\newcommand {\sqt} {\sqrt{t}}
\newcommand {\reg} {\frac{1}{\epsilon}}
\newcommand {\fpisq} {(4\pi)^2}
\newcommand {\Lcal}{{\cal L}}
\newcommand {\Ocal}{{\cal O}}
\newcommand {\Dcal}{{\cal D}}
\newcommand {\Ecal}{{\cal E}}
\newcommand {\Ncal}{{\cal N}}
\newcommand {\Mcal}{{\cal M}}
\newcommand {\Dvec}{{\vec D}}
\newcommand {\dvec}{{\vec d}}
\newcommand {\Evec}{{\vec E}}
\newcommand {\Hvec}{{\vec H}}
\newcommand {\Vvec}{{\vec V}}
\newcommand {\Btil}{{\tilde B}}
\newcommand {\ctil}{{\tilde c}}
\newcommand {\Stil}{{\tilde S}}
\newcommand {\Ztil}{{\tilde Z}}
\newcommand {\altil}{{\tilde \alpha}}
\newcommand {\betil}{{\tilde \beta}}
\newcommand {\latil}{{\tilde \lambda}}
\newcommand {\phitil}{{\tilde \phi}}
\newcommand {\Phitil}{{\tilde \Phi}}
\newcommand {\natil} {{\tilde \nabla}}
\newcommand {\Shat}{{\hat S}}
\newcommand {\xhat}{{\hat x}}
\newcommand {\Zhat}{{\hat Z}}
\newcommand {\Gahat}{{\hat \Gamma}}
\newcommand {\nah} {{\hat \nabla}}
\newcommand {\gh}  {{\hat g}}
\newcommand {\labar}{{\bar \lambda}}
\newcommand {\cbar}{{\bar c}}
\newcommand {\dbar}{{\bar d}}
\newcommand {\bbar}{{\bar b}}
\newcommand {\Bbar}{{\bar B}}
\newcommand {\psibar}{{\bar \psi}}
\newcommand {\chibar}{{\bar \chi}}
\newcommand {\dbartil}{{\tilde {\bar d}}}
\newcommand {\bbartil}{{\tilde {\bar b}}}
\newcommand {\intfx} {{\int d^4x}}
\newcommand {\change} {\leftrightarrow}
\newcommand {\ra} {\rightarrow}
\newcommand {\larrow} {\leftarrow}
\newcommand {\ul}   {\underline}
\newcommand {\pr}   {{\quad .}}
\newcommand {\com}  {{\quad ,}}
\newcommand {\q}    {\quad}
\newcommand {\qq}   {\quad\quad}
\newcommand {\qqq}   {\quad\quad\quad}
\newcommand {\qqqq}   {\quad\quad\quad\quad}
\newcommand {\qqqqq}   {\quad\quad\quad\quad\quad}
\newcommand {\qqqqqq}   {\quad\quad\quad\quad\quad\quad}
\newcommand {\qqqqqqq}   {\quad\quad\quad\quad\quad\quad\quad}
\newcommand {\lb}    {\linebreak}
\newcommand {\nl}    {\newline}

\newcommand {\vs}[1]  { \vspace*{#1 cm} }

\newcommand {\MPL}  {Mod.Phys.Lett.}
\newcommand {\NP}   {Nucl.Phys.}
\newcommand {\PL}   {Phys.Lett.}
\newcommand {\PR}   {Phys.Rev.}
\newcommand {\PRL}   {Phys.Rev.Lett.}
\newcommand {\CMP}  {Commun.Math.Phys.}
\newcommand {\AP}   {Ann.of Phys.}
\newcommand {\PTP}  {Prog.Theor.Phys.}
\newcommand {\NC}   {Nuovo Cim.}
\newcommand {\CQG}  {Class.Quantum.Grav.}


\font\smallr=cmr5
\def\ocirc#1{#1^{^{{\hbox{\smallr\llap{o}}}}}}
\def\ogamma{\ocirc{\gamma}{}}
\def\oM{{\buildrel {\hbox{\smallr{o}}} \over M}}
\def\osigma{\ocirc{\sigma}{}}

\def\overleftrightarrow#1{\vbox{\ialign{##\crcr
 $\leftrightarrow$\crcr\noalign{\kern-1pt\nointerlineskip}
 $\hfil\displaystyle{#1}\hfil$\crcr}}}
\def\overnab{{\overleftrightarrow\nabslash}}

\def\va{{a}}
\def\vb{{b}}
\def\vc{{c}}
\def\tilpsi{{\tilde\psi}}
\def\tbpsi{{\tilde{\bar\psi}}}

\def\Dslash{{}\hbox{\hskip2pt\vtop
 {\baselineskip23pt\hbox{}\vskip-24pt\hbox{/}}
 \hskip-11.5pt $D$}}
\def\nabslash{{}\hbox{\hskip2pt\vtop
 {\baselineskip23pt\hbox{}\vskip-24pt\hbox{/}}
 \hskip-11.5pt $\nabla$}}
\def\xislash{{}\hbox{\hskip2pt\vtop
 {\baselineskip23pt\hbox{}\vskip-24pt\hbox{/}}
 \hskip-11.5pt $\xi$}}
\def\leftnabla{{\overleftarrow\nabla}}

\def\delL{{\delta_{LL}}}
\def\delG{{\delta_{G}}}
\def\delc{{\delta_{cov}}}


\begin{flushright}
US-98-01\\
hep-th/9802043
\end{flushright}
\vs 1

\begin{center}
{\Large\bf Conformal Anomaly 
in  4D Gravity-Matter Theories Non-minimally  
Coupled with Dilaton }

\vspace{2cm}

{\large Shoichi ICHINOSE
\footnote{
e-mail : ichinose@u-shizuoka-ken.ac.jp}
and Sergei D. ODINTSOV$^{\dag}$
\footnote{
e-mail : odintsov@tspi.tomsk.su }
}
\vspace{1cm}

{\large Department of Physics, 
Universuty of Shizuoka \\
Yada 52-1, Shizuoka 422-8526, JAPAN\\}
{\large $^{\dag}$Tomsk Pedagogical University,
634041 Tomsk, RUSSIA }

\end{center}
\vfill
{\large Abstract}\nl
The conformal anomaly for 4D 
 gravity-matter theories, which are  non-minimally coupled with
the dilaton, is systematically studied.
Special care is taken for:
rescaling of fields, treatment of total derivatives, 
hermiticity of the system operator and choice of measure.
Scalar, spinor and vector fields
are taken as the matter quantum fields and 
their explicit conformal anomalies 
in the gravity-dilaton background are
found. 
The cohomology analysis is done and some new conformal invariants
and trivial terms, involving the dilaton, are obtained. The symmetry
of the constant shift of the dilaton field plays an important role.
The general structure of the 
conformal anomaly is examined. 
It is shown that the dilaton affects the conformal anomaly
characteristically for each case:\ 
1)[Scalar]\ The dilaton changes the conformal anomaly only
by a new conformal invariant, $I_4$;
2)[Spinor]\ The dilaton does {\it not} change the conformal anomaly;
3)[Vector]\ The dilaton changes the conformal
anomaly by three new (generalized) conformal invariants, $I_4,I_2,I_{1}$.
We present some
new anomaly formulae which are useful for practical
calculations. Finally, the anomaly induced action is calculated
for the dilatonic Wess-Zumino model. We point out that 
the coefficient of the total derivative term in the conformal 
anomaly for the 2D scalar coupled to a dilaton is ambiguous. 
This resolves 
the disagreement between calculations in refs.\cite{ENO,NO,SI97,KLV} 
and the result of Hawking-Bousso\cite{BH}.
\vs 1
\begin{itemize}
\item PACS NO:\ 04.62.+v,11.25.Db,11.30.-j,14.80.-j
\item Keywords:\ conformal anomaly, dilaton gravity, dilatonic Wess-Zumino
model, 
anomaly formula
\end{itemize}
\vs 2
\section{Introduction}
Stimulated by string theory, the higher dimensional
approach to the 4D theory becomes gradually
important. Indeed it provides fruitful possibilities for 
theoretical analysis. Kaluza-Klein type dimensional
reduction is the most popular one. 
However realizations of such reductions 
depend on the chosen higher dimension, the method 
of reduction, the choice of a symmetry ansatz, and so on. At present
there seems to be no fixed rule to find the "right" one. 
We first note
 that, in  low-energy effective actions, 
some characteristic fields, 
such as the dilaton and the anti-symmetric tensor commonly appear
as remnants of the extra dimensions. 
Through the study of the dynamical effect  
of such fields,
we can learn some common 
features of 4D effective theories that the higher-dimensional
theories produce in the low energy limit. 

We focus on the dilaton field for simplicity. 
Its effect on the Weyl anomaly is examined. We start from the Weyl
invariant action in order to set up the problem clearly.
The coupling of the dilaton is fixed by 
keeping the Weyl symmetry. 
The most characteristic aspect is its  
 exponential form. 
Furthermore it couples to most fields just like the
gravitational field itself. ( From the standpoint of the string
theory or Kaluza-Klein theory, 
this is a natural aspect because the dilaton 
comes from a gravitational component in the compactified
dimensions.) 
We treat the dilaton as the background field in the same
way as the gravitational field. We take the standpoint
that the quantum treatment of the dilaton, like
the gravitational field, should be explained by
the string theory.

It is interesting that the
dilaton is  expected to play a role of a coupling in string theory. 
Indeed, in the present treatment,  
the dilaton
is invariant under all symmetry transformations 
and behaves just
like a coupling constant except that it depends on the space coordinates.
The dilaton transforms as a scalar under the general coordinate
transformation. Under the Weyl transformation, it does not change
\bea
\label{int.1}
\p'=\p\pr
\eea
In this point the dilaton transforms differently from the scalar 
field except in the 2D  space (see Appendix C),
which is natural because of the dimensional difference between the dilaton
and a scalar field.

As for the role of the dilaton, some 
 interesting arguments have been presented recently. 
Even at the classical level,
the dilaton can make the singularity
behaviour weaker in the early universe cosmology\cite{SR97}.
In 2D or 4D black hole physics, its presence influences
the Hawking radiation \cite{BH,ENO,NO,SI97,KLV}.
As for a recent development using the conformal anomaly, we note
ref.\cite{EO97} where a nonperturbative approach to
the conformal field theory in higher dimensions is presented (
for early works in the study of higher dimensional 
conformal field theory, see\cite{C}).
The conformal anomaly
in 4D dilaton gravity-vector  has been recently found in 
 \cite{NO98} where the induced action is examined and by
\cite{SI98} where the general structure of the conformal anomaly
is analysed.

The above is the general background to the present work.
In the following we 
will concretely analyse the following points.
\begin{enumerate}
\item
We take the spin=0,1/2,1 matter fields on dilatonic curved
backgrounds and examine the conformal
anomaly due to quantum effects. 
The dilaton's presence is a new point here.
\item
We  are interested in the
general structure of the conformal anomaly.
\item
In the evaluation of the anomaly we pay attention to the following
points:\ the arbitrariness of 
total derivatives or partial derivatives in the operators, 
rescaling of fields and the choice of measure.
\item
We present some useful anomaly formulae and show how to use them.
\item
We check the previous work, which is obtained as a special case 
:\ dilaton field = const by using a different approach (heat-kernel
regularization and Fujikawa method).
\end{enumerate}

We begin with the explanation of the present formalism and the cohomology
analysis in Sec.2. 
In Sec.3 the scalar theory is considered. This is the simplest
case and is the best model in which the present analysis is 
explained. 
In Sec.4 the vector
theory is considered. We must take into account the gauge fixing
which introduces some complication. It breaks the Weyl symmetry
 as well as the gauge symmetry. 
The final result is therefore highly non-trivial.
In Sec.5 we treat the spinor theory. It turns out that the dilaton
does not affect the spinor theory.
Finally we discuss the anomaly induced effective action in Sec.6. 
As an example
the dilatonic Wess-Zumino model is considered.
We conclude in Sec.7. 
In appendix A and B, we derive useful {\it anomaly formulae}
used in the text. We use  heat-kernel
regularization for the ultra-violet divergences 
and follow the Fujikawa method\cite{KF79}. 
The results turn out to be the same as the {\it counterterm
formulae}  already
obtained by other authors, where  dimensional 
regularization was used
and the (log-)divergent part of the effective
action was calculated.
In the present approach the anomaly
term is definitely determined (no ambiguity of the total
divergence)  
once the system operator (elliptic differential
operator) and the regularization are fixed. In appendix C the
Weyl transformations of various fields are listed. In appendix D, 
the list of "trivial terms" is presented. In appendix E we discuss the  
conformal anomaly for a two-dimensional dilaton coupled scalar. 

\section{Formalism and Cohomology Analysis}
We consider the system of quantum matter
(scalar,vector,spinor,etc.) fields $f(x)$, in the background
metric $g_\mn(x)$ and the background dilaton field $\p(x)$, described by the
action $S[f;g,\p]$. The (Wilsonian) effective action $\Ga[g,\p]$ is defined 
by
\begin{eqnarray}
\e^{\Ga[g,\p]}=\int\Dcal f~exp\{\, S[f;g,\p]\,\}\pr  \label{f.1}
\end{eqnarray}
Under the Weyl transformation, the classical lagrangian 
is assumed to be
invariant.
\bea
\label{f.2}
\p'=\p\com\q 
{g_\mn}'=\e^{2\al(x)}g_\mn\com\q
f'=\e^{c\al(x)}f\com \nn
S[f';g',\p'=\p]=S[f;g,\p]\com
\eea
where $c$ is some constant (the scale dimension of the matter field).
The effective action $\Ga$ is generally changed
due to the measure (quantum) effect which is the origin of
the conformal anomaly\cite{KF79}. The change is given by
\bea
\label{f.3}
\left.\frac{\del\Ga}{\del\al(x)}\right|_{\al=0}
=2g_\mn(x)\frac{\del\Ga}{\del g_\mn(x)}
\equiv 2g_\mn <T^\mn >\com 
\eea
where $T^\mn$ is the energy-momentum tensor. This quantity is
the conformal (Weyl) anomaly and is evaluated as follows.
\begin{eqnarray}
\e^{\Ga[g',\p'=\p]}=\int\Dcal f'~exp\{\, S[f';g',\p'=\p]\, \}\nn
=\int\Dcal f~\det(\frac{\pl f'}{\pl f}) exp\{\, S[f;g,\p]~\}
\pr  \label{f.4}
\end{eqnarray}
We can evaluate the Jacobian in the heat-kernel regularization.
\begin{eqnarray}
\ln~\det~\frac{\pl f'}{\pl f}=\Tr~\ln \{ \e^{c\al(x)}\del^4(x-y)\}
=c\,\Tr\{ \al(x)\del^4(x-y)\}+O(\al^2)\nn
=c\lim_{t\ra +0}\Tr\left\{\al(x)<x|\e^{-t\Dvec}|y>\right \}+O(\al^2)\com\nn
g_\mn <T^\mn>=\frac{c}{2}\tr\left\{<x|\e^{-t\Dvec}|x>\right \}|_{t^0}
\pr  \label{f.5}
\end{eqnarray}
By differentiating
the equation (\ref{f.4}) w.r.t.(with respect to) the background sources $g_\mn$ and $\p$,
we can obtain various Ward-Takahashi identites which take into account
the conformal anomaly. They correspond to 1-loop graphs of 2-point,
3-point, ... functions of corresponding currents.
Taking $t^0$ part in (\ref{f.5}) 
corresponds to considering the log-divergent part in the dimensional
regularization. All these things and the fact eq.(\ref{f.5}) correctly gives
the all known anomalies (both chiral and Weyl) 
were confirmed in \cite{II96}.
Our task therefore reduces to the evaluation of\  
$\tr\{<x|\e^{-t\Dvec}|x>\}|_{t^0}$.

It is well-known the anomaly can be fixed to some extent
by the cohomology of the corresponding symmetry operators.
It is powerful in that no explicit calculation is required 
(hence we need not to worry about the regularization ambiguity) and only
some consistency relations due to the symmetries are essential. 
Especially for the chiral
anomaly, the anomaly is fixed except one overall coefficient.
Let us do the cohomology analysis for the present conformal
anomaly. 

Following Ref.\cite{BR83}
we consider the infinitesimal Weyl transformation in (\ref{f.2}),
$\del g_\mn=2\al(x)g_\mn, \del\p=0$ and regard the infinitesimal
parameter $\al(x)$ as an anti-commuting one. 
\footnote{
Only from this paragraph to the end of this section, 
the Weyl transformation parameter $\al(x)$
is  infinitesimal and anti-commuting.
                }
Then we can introduce
the coboundary operator $\Si$ which acts on the functional space
of $g_\mn$ and $\p$. 
\bea
\label{f.6}
\Si=\intfx 2\al(x) g_\mn(x)\frac{\del}{\del g_\mn(x)}\com\q
\Si^2=0\com
\eea
which is a nilpotent operator. It acts on the effective action $\Ga[g,\p]$ as
\bea
\label{f.7}
\Si\Ga=\hbar\Del+O(\hbar^2)\com
\eea
where the right hand side is expanded w.r.t. the loop-number($\hbar$).
In the anomaly, we usually consider only 1-loop part. 
$\Del$, after removing "trivial terms"(see Appendix D), is defined as the anomaly
in the cohomology.
Because of the
nilpotency property, $\Si^2=0$, the term $\Del$ must satisfy the
following consistency condition.
\bea
\label{f.8}
\Si\Del=0\pr
\eea
We impose the following conditions on the "basis" of the function space
of $\Del$.
\begin{enumerate}
\item 
4D general coordinate invariance. 
\item
Physical dimension is 4 ( Number of derivatives is 4).
\item
"Ghost" number is 1. (This means we consider those terms which contain
one $\al$ only.) 
\item
Symmetry of constant shift:\ 
$P(\phi) \rightarrow P(\phi)+\mbox{const}$ . 
\end{enumerate}
The last item 4 is the speciality of the dilaton field. 
(Compare with the scalar field considered in \cite{BR83}.)
For the familiar case $P(\phi)=\phi$, the transformation
is the constant shift of the dilaton.
The actions involving the dilaton (see
(\ref{s2.1}), (\ref{s1}) and (\ref{sp.1})  in the following text)
are essentially invariant because
the transformation  changes the actions
only by an overall constant factor,
which can be absorbed by the constant rescaling of the
matter fields:\  
$f  \rightarrow  f \times \mbox{const}$.
This symmetry is very important because it
forbids, as the anomaly terms,  
the appearance of the terms like 
$\sqrt{g}(P_{,\mu}P^{,\mu})^2P^2,\sqrt{g}(P_{,\mu}P^{,\mu})^2P^4,
\sqrt{g}C_{\mu\nu\lambda\sigma}C^{\mu\nu\lambda\sigma}P^2,\cdots$.

Let us see how the consistency condition (\ref{f.8}) restricts $\Del$. In Table 1
we list all the possible terms, $\Del_i$, which satisfy 
the above four conditions and
$\Si$-transformed ones for each, $\Si\Del_i$.　
Before the consistency condition is imposed,  
the most general expression is given as
\bea
\label{f.9}
\Del=\sum_i c_i\Del_i\pr
\eea
From the consistency condition we obtain the following relations between
coefficients.
\bea
\label{f.10}
\mbox{i)}\q c_1+c_2+c_3=0\pr\nn
\mbox{ii)}\q 6c_6+c_7=0\com\q -2c_7-4c_9=0\com\q -2c_7+4c_8+2c_9=0\pr\nn
\mbox{iii)}\q 2c_{14}-c_{15}=0\pr\nn
\mbox{iv)}c_{16}+3c_{19}=0\com\q 2c_{17}+c_{18}=0\pr
\eea
Other coefficients are arbitrary.\nl
\vs 1
\begin{tabular}{|c|c|c|}
\hline
No  & $\Del_i$                                 & $\Si\Del_i$                            \\
\hline
\hline
1 & $\int dV R_{\mn\ls}R^{\mn\ls}\al(x)$    &   $4 \int dV R \na^2\al\cdot\al $   \\
\hline
2 & $\int dV R_{\mn}R^{\mn}\al(x)$    &   $4 \int dV R \na^2\al\cdot\al $   \\
\hline
3 & $\int dV R^2\al(x)$                &   $12 \int dV R \na^2 \al\cdot\al$   \\
\hline
4 & $\int dV (\na^2R)\cdot\al(x)$             &   $0 $   \\
\hline
\hline
5 & $\int dV (P_{,\m}P^{,\m})^2\al(x)$   &   $0 $   \\
\hline
\hline
6 & $\int dV RP_{,\m}P^{,\m}\al(x)$   & $ 6\int dV P_{,\m}P^{,\m}\na^2\al\cdot\al $   \\
\hline
7 & $\int dV R^\mn P_{,\m}P_{,\n}\al(x)$   & $ \int dV (P_{,\m}P^{,\m}\na^2\al
                            -2P^{,\mn}P_{,\m}\al_{,\n}-2\na^2P\cdot P^{,\m}\al_{,\m})\al $   \\
\hline
8 & $\int dV (\na^2 P)^2\cdot\al(x)$   & $ 4\int dV \na^2P\cdot P^{,\m}\al_{,\m}\al $   \\
\hline
9 & $\int dV P_{,\mn}P^{,\mn}\al(x)$   & $ \int dV (
                            -4P^{,\mn}P_{,\m}\al_{,\n}\al+2\na^2P\cdot P^{,\m}\al_{,\m}\al) $   \\
\hline
10 & $\int dV \na_\m(\na^2P\cdot P^{,\m})\cdot\al(x)$   & $ 0 $   \\
\hline
11 & $\int dV \na_\m(P^{,\mn}P_{,\n})\cdot\al(x)$   & $ 0 $   \\
\hline
12 & $\int dV \na^2(P_{,\m}P^{,\m})\cdot\al(x)$   & $ 0 $   \\
\hline
13 & $\int dV \na_\m\na_\n(P^{,\m}P^{,\n})\cdot\al(x)$   & $ 0 $   \\
\hline
\hline
14& $\int dV P_{,\m}P^{,\m}(\na^2P)\cdot\al(x)$ & $ 2\int dV P_{,\m}P^{,\m}P_{,\n}
                                                    \al^{,\n}\cdot\al $   \\
\hline
15& $\int dV P^{,\m}P^{,\n}P_{,\mn}\al(x)$ &   $ -\int dV P_{,\m}P^{,\m}P_{,\n}
                                                                 \al^{,\n}\cdot\al $   \\
\hline
\hline
16& $\int dV \na^4P\cdot\al(x)$   & $-2\int dV P_{,\m}\na^2\al
                                                               \cdot\al^{,\m} $ \\
\hline
17& $\int dV R\na^2P\cdot\al(x)$      &   $\int dV (6\na^2P\cdot\na^2\al
                                                  +2RP_{,\m}\al^{,\m})\cdot\al $   \\
\hline
18& $\int dV R^\mn P_{,\mn}\cdot\al(x)$      &   $\int dV (3\na^2P\cdot\na^2\al
                                                  +RP_{,\m}\al^{,\m})\cdot\al $   \\
\hline
19& $\int dV \na_\m(R P^{,\m})\cdot\al(x)$      &   $-6\int dV
                                              P^{,\m}\na^2\al\cdot \al_{,\m} $   \\
\hline
\multicolumn{3}{c}{\q}                                                 \\
\multicolumn{3}{c}{Table 1\ \ List of all possible terms $\Del_i$ which satisfy the four conditions in Sec.2  }\\
\multicolumn{3}{c}{ and their $\Si$-transformed ones(see eq.(\ref{f.6})). $\int dV\equiv \intfx\sqg$.        }\\
\end{tabular}

Since we have 7 relations among 19 terms, 19$-$7=12 terms remain as the possible
Weyl anomaly terms appearing in the gravity-dilaton background system. Among
the 12 terms, there are "trivial" terms which can be freely adjusted by
introducing local conter-terms in the action. 
In order to find all trivial terms, we list, in Table 2, 
all the possible local counterterms $\ga_i$
and the $\Si$-transformed ones for each, $\Si\ga_i$.
The symmetries which define the counterterms are the same as
those for $\Del$ except the condition 3.\ "Ghost" number is 0. \nl
\vs 1

\begin{tabular}{|c|c|c|}
\hline
No  & $\ga_i$                                 & $\Si\ga_i$                            \\
\hline
\hline
1 & $\int dV R_{\mn\ls}R^{\mn\ls}$    &   $4 \int dV \al \na^2 R $   \\
\hline
2 & $\int dV R_{\mn}R^{\mn}$    &   $4 \int dV \al \na^2 R $   \\
\hline
3 & $\int dV R^2$                &   $12 \int dV \al \na^2 R$   \\
\hline
4 & $\int dV (\na^2R)$             &   $0 $   \\
\hline
\hline
5 & $\int dV (P_{,\m}P^{,\m})^2$   &   $0 $   \\
\hline
\hline
6 & $\int dV RP_{,\m}P^{,\m}$   & $ 6\int dV \al\na^2(P_{,\m}P^{,\m}) $   \\
\hline
7 & $\int dV R^\mn P_{,\m}P_{,\n}$   & $ \int dV \al \{ \na^2(P_{,\m}P^{,\m})
                             +2\na_\m\na_\n(P^{,\m}P^{,\n}) \}$   \\
\hline
8 & $\int dV (\na^2 P)^2$   & $ -4\int dV \al\na_\m (\na^2P\cdot P^{,\m}) $   \\
\hline
9 & $\int dV P_{,\mn}P^{,\mn}$   & $ 2\int dV \al\na_\m( 2P^{,\mn}P_{,\n}
                                                          -\na^2P\cdot P^{,\m})  $   \\
\hline
10 & $\int dV \na_\m(\na^2P\cdot P^{,\m})$   & $ 0 $   \\
\hline
11 & $\int dV \na_\m(P^{,\mn}P_{,\n})$   & $ 0 $   \\
\hline
12 & $\int dV \na^2(P_{,\m}P^{,\m})$   & $ 0 $   \\
\hline
13 & $\int dV \na_\m\na_\n(P^{,\m}P^{,\n})$   & $ 0 $   \\
\hline
\hline
14& $\int dV P_{,\m}P^{,\m}(\na^2P)$ & $ -2\int dV \al \na_\n(P_{,\m}P^{,\m}P^{,\n})
                                                                           $   \\
\hline
15& $\int dV P^{,\m}P^{,\n}P_{,\mn}$ &   $ \int dV \al \na_\n(P_{,\m}P^{,\m}P^{,\n})
                                                                  $   \\
\hline
\hline
16& $\int dV \na^4P$             &   $0 $   \\
\hline
17& $\int dV R\na^2P$      &   $2\int dV \al \{3\na^4P-\na^\m(RP_{,\m})\} $   \\
\hline
18& $\int dV R^\mn P_{,\mn}$      &   $\int dV \al \{3\na^4P-\na^\m(RP_{,\m})\} $   \\
\hline
19& $\int dV \na_\m(R P^{,\m})$      &   $0 $   \\
\hline
\multicolumn{3}{c}{\q}                                                 \\
\multicolumn{3}{c}{Table 2\ \ List of all possible local counterterms $\ga_i$  }\\
\multicolumn{3}{c}{ and their $\Si$-transformed ones. $\int dV\equiv \intfx\sqg$.        }\\
\end{tabular}

Therefore we conclude, from the cohomology analysis, the Weyl anomaly 
in the gravity-dilaton background system is composed of
four invariants $(I_0, I_4, I_2, I_1)$, the Euler term $\sqg E$, seven
trivial terms $(J_0,J_{2a},J_{2b},J_{2c},J_{2d},J_3,J_1)$.
\footnote{
$I_0\sim\Del_1-2\Del_2+(1/3)\Del_3,\ \sqg E\sim\Del_1-4\Del_2+\Del_3,\ 
J_0\sim\Del_4,\ I_4\sim\Del_5,\ I_2\sim -(1/6)\Del_6+\Del_7+(3/4)\Del_8-(1/2)\Del_9,\ 
I_1\sim -(1/2)\Del_{17}+\Del_{18},\ J_{2a}\sim\Del_{12},\ J_{2b}\sim\Del_{11}-\Del_{10},\ 
J_{2c}\sim\Del_{13},\ J_{2d}\sim\Del_{11},\ J_3\sim\Del_{14}+2\Del_{15},\ 
J_1\sim 3\Del_{16}-\Del_{19}.$
}
\bea
\label{f.11}
\frac{\del}{\del\al(x)}\Del=b_0I_0+b_E\sqg E+c_0J_0+b_4I_4+b_2I_2+b_1I_1\nn
+c_{2a}J_{2a}+c_{2b}J_{2b}+c_{2c}J_{2c}+c_{2d}J_{2d}+c_3J_3+c_1J_1
\pr
\eea
The definite expressions for the symbols $(I_i,E,J_i)$ appear in the following text 
and in Appendix D.

In the cohomology approach, there generally remains not a few
undetermined coefficients of conformal anomaly terms, compared with the chiral anomaly. 
In order to
fix them completely, the "full-fledged" dynamical calculation
like the present approach is indispensable.

\section{Trace anomaly for 4D dilaton coupled scalar}
Let us start with the Lagrangian
\bea
\label{s2.1}
S_1[\vp;g,\p]=\half\intfx\sqg \e^{P(\p)}\vp (-\na^2+\xi R)\vp\com
\eea
where $\vp,g_\mn$ and $\p$ are the scalar, graviton and dilaton
fields, respectively.
The form of $P(\p)$ is often chosen to be 
$\mbox{const}\times \p$, but we will 
keep a more general form. 
We treat $g_\mn$ and $\p$ as the background fields and
the scalar matter field, $\vp$, as the quantum field.
At the value of $\xi=-\sixth$, (\ref{s2.1})
is exactly (not ignoring total derivatives) conformally
invariant.
\footnote{
We cannot take the following actions
(compare it with the 2D case\cite{SI97}):
\bea
S_2[\vp;g,\p]=\half\intfx\sqg \e^{P(\p)}(\na_\m\vp\cdot\na^\m\vp
+\xi R\vp^2)\com\nn
{S_2}'[\vp;g,\p]=
S_2[\vp;g,\p]-\half\intfx\sqg \na_\m(\e^{P(\p)}\vp\na^\m\vp)\nn
=\half\intfx\sqg \vp\e^{P(\p)}(-\na^2-P_{,\m}\na^\m+\xi R)\vp
\pr
\eea
These two actions, in the $P(\p)=\mbox{const}$ case, 
differ from (\ref{s2.1}) only 
by a total derivative.
For the general case, however, they
are not conformally invariant even for $\xi=-\sixth$.
The dilaton can select the coupling with 
the ordinary matter fields among some possibilities which differ 
only by the total derivative term (surface term)
at the $P(\p)=\mbox{const}$ limit. 
Note that 
the operator from ${S_2}'$, 
$\Dvec_2\equiv\e^{P(\p)}(-\na^2-P_{,\m}\na^\m+\xi R)$,
is hermitian ($\Dvec_2={\Dvec_2}^\dag$) for the natural definition:\  
$\intfx\sqg(\Dvec_2\vp)^\dag\vp=\intfx\sqg\vp^\dag{\Dvec_2}^\dag\vp$.
}
\bea
\label{s2.2}
\p'=\p\com\q {g_\mn}'=\e^{2\al(x)}g_\mn\com\q
\vp'=\e^{-\al(x)}\vp\pr
\eea

In order to normalize the kinetic term of the scalar field, 
we rescale it as
\footnote{
A way to avoid the rescaling (\ref{s2.5}) is to
take fully the dilaton coupling into the operator:\ 
$\Dvec_1=\e^{P(\p)} (-\na^2+\xi R)$. This is, however, 
not hermitian($\Dvec_1\neq(\Dvec_1)^\dag$) for the natural 
definition:\ 
$\intfx\sqg(\Dvec_1\vp)^\dag\vp=\intfx\sqg\vp^\dag(\Dvec_1)^\dag\vp$.
It should be compared with $\Dvec_2$ in the previous footnote.
}
\bea
\label{s2.5}
\Phi=\e^{\half P(\p)}\vp\pr
\eea
Adding a total derivative term, $S_1$ changes to
\bea
\label{s2.6}
S=\half\intfx\sqg\Phi\{-\na^2-\fourth P_{,\m}P^{,\m}
+\xi R\}\Phi
=S_1-\fourth \intfx\sqg\na_\m(\Phi^2P^{,\m})
\pr
\eea
We do the second rescaling in order to make
the integration measure BRST-invariant\cite{KF83} 
with respect to (w.r.t.)
4D general coordinate symmetry.
\bea
\label{s2.7}
{\tilde \Phi}=\fg \Phi
\pr
\eea
We can write $S$ as
\bea
\label{s2.8}
 S=\half\intfx{\tilde \Phi}\Dvec{\tilde \Phi}\ ,\ 
\Dvec\equiv\fg(-\na^2-{\bar \Mcal})\invfg\ ,\ 
{\bar \Mcal} =-\x R+\fourth P_{,\m}P^{,\m}\ .
\eea

We generally have the freedom of the choice of 
surface terms (like (\ref{s2.6})) and of
rescaling the variables ( like (\ref{s2.5}) and 
(\ref{s2.7}) ).
We cannot keep $S_1$ as the action, because the corresponding
operator 
$\Dvec'=\fg(-\na^2-{\bar \Mcal}+\half \na^2P+P^{,\m}\na_\m)\invfg$, 
is not hermitian due to the presence of the first derivative term.
We can define the functional space ${\cal F}$, 
in terms of the eigen functions of $\Dvec$, as\ :  
\bea
\label{s2.8b}
& \Dvec u_n=\la_nu_n\com\q
{\cal F}=\{ {\tilde \Phi}=\sum_n c_nu_n|c_n\in \mbox{Complex},
n=0,1,2,\cdots\}\pr &
\eea
The inner product between functions and
the hermite conjugate of the operator $\Dvec$ are defined as
\bea
\label{s2.8c}
& <\Phitil_1|\Phitil_2>\equiv\intfx\Phitil_1^\dag\Phitil_2
=\intfx\sqg\Phi_1^\dag\Phi_2=\intfx\sqg \e^{P(\p)}
\vp_1^\dag\vp_2\com  & \nn
& <\Dvec\Phitil_1|\Phitil_2>\equiv <\Phitil_1|\Dvec^\dag\Phitil_2>\pr &
\eea
Under this definition, the hermiticity
of the operator $\Dvec$ in (\ref{s2.8}) holds true. 
Taking the eigen functions
$\{u_n\}$ orthonormal, the integration measure is given by
\bea
\label{s2.8d}
<u_n|u_m>=\intfx~ u_n^\dag u_m=\del_{nm}\com\q
\prod_n dc_n\equiv\Dcal\Phitil=\Dcal(g^{1/4}\e^{\half P(\p)}\vp)\pr
\eea
This rather standard procedure is important especially in the dilaton
coupled system in order to precisely fix the integration measure\cite{SI97}.
We note that the requirement of the hermiticity of 
the system operator $\Dvec$ plays an important role here.

In Appendix A, the anomaly formula for $\Dvec=\fg(-\na^2-\Mcal)\invfg$
with arbitrary background scalar $\Mcal$ is obtained
in the present approach (the heat-kernel regularization and Fujikawa
method):
\bea
\label{s2.9}
\mbox{Anomaly}=\frac{\sqg}{\fpisq}\Tr \{ \sixth\na^2 (\Mcal-\sixth R)
+\half (\Mcal-\sixth R)^2\nn
+\frac{1}{180}(R_{\mn\ls}R^{\mn\ls}-R_\mn R^\mn
+0\times R^2-\na^2 R)\cdot {\bf 1}
                                       \}\com
\eea
where "$\Tr$" means "trace" over the field suffix $i$:\ 
$\Tr \Mcal=\sum_{i=1}^{N_s}\Mcal_{ii}$ and $({\bf 1})_{ij}=\del_{ij}$. 
Formula (\ref{s2.9}) ( and the vector version (\ref{s4}) appearing
later ) is the same as the counterterm formula given in 
\cite{BV81,BV85,BOS92} where the dimensional regularization was used.
Taking $N_s=1$
for the present case we find
\bea
\label{s2.10}
\mbox{Anomaly}|_{\xi}=\frac{\sqg}{\fpisq}\ \{ 
\frac{1}{32}(P_{,\m}P^{,\m})^2    
+\frac{1}{180}(R_{\mn\ls}R^{\mn\ls}-R_\mn R^\mn-\na^2 R)\nn
+(\xi+\sixth)\left(-\sixth\na^2R+\half(\xi+\sixth)R^2
-\fourth RP_{,\m}P^{,\m}
              \right)                   
+\frac{1}{24}\na^2(P_{,\m}P^{,\m})
                                       \}\pr
\eea
This result is consistent
with \cite{NO6143} where  
dimensional regularization was used.
The above result is not a conformal anomaly in the strict sense
because $\Dvec$ of (\ref{s2.8}) is not conformal invariant
for the general value of $\xi$. 
At the classical conformal limit $\xi=-\sixth$, the above
result reduces to the "true" conformal anomaly:
\bea
\label{s2.11}
\mbox{Anomaly}|_{\xi=-1/6}=\frac{\sqg}{\fpisq}\ \{ 
\frac{1}{32}(P_{,\m}P^{,\m})^2    \nn
+\frac{1}{360}(3C_{\mn\ls}C^{\mn\ls}-E)
-\frac{1}{180}\na^2R
+\frac{1}{24}\na^2(P_{,\m}P^{,\m})
                                       \}\com
\eea
where $C_{\mn\ls}$ and $E$ are the conformal tensor(Appendix C) 
and the
Euler term, respectively, which are defined by
\bea
\label{s2.12}
C_{\la\mn\si}=R_{\la\mn\si}+\half(g_{\la\n}R_{\m\si}
-g_\ls R_\mn+g_{\m\si}R_{\la\n}-g_\mn R_\ls)\nn
+\sixth (g_\ls g_\mn-g_{\la\n}g_{\m\si})R\com\nn
C_{\la\mn\si}C^{\la\mn\si}=-2R_\mn R^\mn
+R_{\mn\ls}R^{\mn\ls}+\frac{1}{3}R^2\com\nn
E=\fourth R_{\mn\ab}R_{\ls\ga\del}\ep^{\mn\ls}\ep^{\ab\ga\del}
=R^2+R_{\mn\ab}R^{\mn\ab}-4R_\mn R^\mn\pr
\eea

In (\ref{s2.11}) we notice two conformal invariants:
\bea
\label{s2.12b}
I_0\equiv\sqg C_{\mn\ls}C^{\mn\ls},
I_4\equiv\sqg(P_{,\m}P^{,\m})^2
                                       \com
\eea
and two "trivial" terms (Appendix D) $J_0$ and $J_{2a}$:
\bea
\label{s2.12c}
J_0\equiv\sqg\na^2R\com\q
K_0\equiv\sqg R^2\com\q
g_\mn\frac{\del}{\del g_\mn (x)}\intfx K_0=6 J_0\com\nn
J_{2a}\equiv\sqg\na^2(P_{,\m}P^{,\m})\ ,\ 
K_{2a}\equiv\sqg R P_{,\m}P^{,\m}\ ,\ 
g_\mn\frac{\del}{\del g_\mn (x)}\intfx K_{2a}=+3 J_{2a}\ .
\eea
The numbers of lower suffixes in $I_0,I_4,J_0,K_0,$ etc, show
the power numbers of $P$ ( see Appendix D).
We note that $I_4$ and $J_{2a}$ are caused by the presence of the dilaton.
Two terms $J_0$ and $J_{2a}$
in (\ref{s2.11}) are arbitrarily adjusted by introducing
finite counterterms, $K_0$ and $K_{2a}$ respectively, 
in the original (local gravitational) action. 
This result says the conformal anomaly at $\xi=-\sixth$
consists of two conformal invariants ($I_0,I_4$), 
the Euler term($\sqg E$) and two
trivial terms ($J_0,J_{2a}$). This confirms the 
general structure of the conformal anomaly discussed in
\cite{DDI76,MD77,DS93,SI98}.

 2D case which is qualitatively the same is discussed in appendix E.

\section{Trace anomaly for 4D dilaton coupled vector}

\subsection{Basic Formalism}

In this section we consider the dilaton coupled photon system.
The action is given by

\bea
\label{s1}
S_V[A;g,\phi]=\int d^4x \sqrt{g}(-{1 \over 4}\e^{P(\phi)}F_\mn F^\mn )\ ,\ 
F_\mn=\na_\m A_\n-\na_\n A_\m\ .
\eea
Besides the 4D general coordinate symmetry, 
this theory has the local Weyl symmetry:
\bea
\label{s1a}
g_\mn'=\e^{2\al(x)}g_\mn\com\q 
A_\m:\ \mbox{fixed}\com\q
\p:\ \mbox{fixed}\com
\eea
and it has the Abelian local gauge symmetry:
\bea
\label{s1b}
A_\m'=A_\m+\na_\m\La(x)\com\q
g_\mn:\ \mbox{fixed}\com\q \p:\ \mbox{fixed}\pr
\eea

We take the following gauge-fixing term:
\bea
\label{s2}
S_{gf}[A;g,\phi]=\int d^4x \sqrt{g}(-\half \e^{P(\phi)}(\na_\m A^\m)^2 )\pr
\eea
The corresponding ghost lagrangian is 
\bea
\label{s2a}
S_{gh}[\cbar,c;g,\phi]=\intfx\sqg \e^{P(\p)}\cbar\na_\m \na^\m c\com
\eea
where $\cbar$ and $c$ are the anti-ghost and ghost fields
respectively. They  are hermitian scalar fields with Fermi statistics
(Grassmannian variables). The total action 
$S[A,\cbar,c;g,\p]=S_V+S_{gf}+S_{gh}$ is invariant under the
BRST symmetry
\bea
\label{s2b}
\del A_\m=\xi\pl_\m c\com\q \del\cbar=\xi\na^\m A_\m\com\q
\del c=0\com\q \del\p=0\com\q \del g_\mn=0\com
\eea
where $\xi$ is the BRST parameter which is a global Grassmannian variable. 
$S_{gf}$\ breaks the Weyl symmetry (\ref{s1a}), besides 
the gauge symmetry (\ref{s1b}).
Therefore we must take close care to define the conformal anomaly
in gauge theory. It is highly non-trivial
whether the general structure suggested by Ref.\cite{MD77,DS93}
holds true in this case.

We assign the following Weyl transformation to $\cbar$ and $c$ 
\cite{KF81}:\ 
\bea
\label{s2c}
\cbar'=\e^{-2\al(x)}\cbar\com\q c'=c\com
\eea
which is chosen in the following way:\ 
i)\ the ghost action (\ref{s2a})
is invariant under global Weyl transformation ($\al(x)$ is
independent of $x$)\ ;\ ii)\ $S_{gf}+S_{gh}$ is invariant, up to
a "BRST-trivial" term, under infinitesimal local Weyl transformation. 

After rescaling the vector and ghost fields 
( for normalizing the kinetic term):\ 
$ \e^{\half P(\p)}A_\m=B_\m\ ,\ \e^{P(\p)}\cbar=\bbar$, 
the actions can be expressed as
\bea
\label{s3}
& S_V+S_{gf}=\int d^4x [
\half \sqg B^\m \{g_\mn \na^2+(\Ncal^\la)_\mn\na_\la+\Mcal_\mn\} B^\n  
+\mbox{total deri.}]\ ,                           & \nn
& S_{gh}=\intfx\sqg\bbar\na^2c\com                & \nn 
& (\Ncal^\la)_\mn=P_{,\m}\del^\la_\n-P_{,\n}\del^\la_\m\com\q
\Mcal_\mn=R_\mn-\fourth g_\mn (P_{,\la}P^{,\la}+2\na^2P)\com &
\eea
Following Fujikawa\cite{KF83}, we take  ( 4D general
coordinate transformation symmetry) 
BRST-invariant measure by taking
$\Btil_\m,\bbartil$ and $\ctil$ instead of $B_\m, \bbar$ and $c$
respectively, as the path-integral variables
\footnote{
The measure is also invariant under the BRST transformation
w.r.t. the Abelian gauge symmetry:
$
\del\Btil_\m=\xi\e^{P/2}g^{-1/8}(\pl_\m\ctil
           -\fourth g^{-1}\pl_\m g\cdot\ctil)\ ,\ 
\del\bbartil=\xi\e^{P/2}g^{1/8}(\na^\m\Btil_\m
-\frac{1}{8}g^{-1}g^{,\m}\Btil_\m-\half P^{,\m}\Btil_\m)\ ,\ 
\del\ctil=0\ ,\ 
\det\{
   \pl({\Btil_\m}',\bbartil',\ctil')/ \pl(\Btil_\n,\bbartil,\ctil)
    \}
=1.	   
$
}
: 
\bea
\label{s3a}
\Btil_\m\equiv g^{1/8} B_\m\ \ (\Btil^\m\equiv g^{3/8} B^\m)\com
\bbartil\equiv\fg\bbar\com\ctil\equiv\fg c\com\nn
S_V+S_{gf}=\int d^4x [
-\half \Btil^\m\Dvec_\m^{~\n} \Btil_\n  +\mbox{total deri.}]\ ,\nn
\Dvec_\m^{~\n}
=-g^{1/8}\{\del_\m^{~\n} \na^2+(\Ncal^\la)_\m^{~\n}\na_\la
+\Mcal_\m^{~\n}\}g^{-1/8}\ ,\nn
S_{gh}=-\intfx \bbartil\ \dvec\ \ctil\com\q 
\dvec\equiv -\fg\na^2\invfg\com\nn
\exp\{\Ga[g,\p]\}=\int\Dcal\Btil\Dcal\bbartil\Dcal\ctil~
\exp (\Stil[\Btil,\bbartil,\ctil;g,\p])\com\nn
\Stil[\Btil,\bbartil,\ctil;g,\p]=S[A,\cbar,c;g,\p]\com
\eea
where $\Ga[g,\p]$ is the (Wilsonian) effective action.

The Weyl transformation of (\ref{s1a}) and that of 
the integration variables
($\Btil_\m'=\e^{\al}\Btil_\m,\ \bbartil'=\bbartil,
\ \ctil'=\e^{2\al}\ctil$) give us the Ward-Takahashi identity
for the Weyl transformation:
\bea
\label{s3b}
\exp\Ga[g',\p]=\int\Dcal\Btil'\Dcal\bbartil\Dcal\ctil'
\exp \Stil[\Btil',\bbartil,\ctil';g',\p]       \nn
=\int\Dcal\Btil\cdot\det\frac{\pl \Btil'}{\pl \Btil}\cdot
\Dcal\bbartil\Dcal\ctil\cdot\det\frac{\pl \ctil'}{\pl \ctil}
\exp \Stil[\Btil',\bbartil,\ctil';g',\p]       \nn
=\int\Dcal\Btil\Dcal\bbartil\Dcal\ctil\cdot
\det(\e^{\al}\del_\m^{~\n}\del^4(x-y))\cdot
\{ \det(\e^{2\al}\del^4(x-y))\}^{-1}         \nn
\times\exp \left[
\Stil[\Btil,\bbartil,\ctil;g,\p]
+\del\Btil_\m\frac{\del\Stil}{\del\Btil_\m}
+\del\ctil\frac{\del\Stil}{\del\ctil}
+\del g_\mn\frac{\del\Stil}{\del g_\mn}
          \right]       \pr
\eea
Considering the infinitesimal transformation
($\del g_\mn=2\al g_\mn,\ \del\Btil_\m=\al\Btil_\m,\ 
\del\ctil=2\al\ctil$) in the above equation, and regularizing
the space delta-functions $\del^4(x-y)$ in terms of the heat-kernels,
we obtain
\bea
\label{s3c}
\Ga[g',\p]-\Ga[g,\p]=2\al g_\mn\frac{\del\Ga}{\del g_\mn}
=\al\left<\Btil_\m\frac{\del\Stil}{\del\Btil_\m}\right>
+2\al\left<\ctil\frac{\del\Stil}{\del\ctil}\right>
+2\al\left<g_\mn\frac{\del\Stil}{\del g_\mn}\right>\nn
+\Tr \ln\{\e^{\al}<x|\e^{-t\Dvec_\m^{~\n}}|y>\}
-\Tr \ln\{\e^{2\al}<x|\e^{-t\dvec}|y>\}\com
\eea
where $t$ is the regularization parameter:\ $t\ra +0$.
The part of variation w.r.t. $\Stil$ gives the "naive"
Ward-Takahashi identity, while the remaining two terms (terms with "Tr")
give the deviation from it. 
(Compare this with the conformal gauge identities in gauge theories 
 on curved space without the dilaton \cite{BOS92}.) 
Therefore 
the last two terms can be regarded as the Weyl anomaly in this theory.
\bea
\label{s3d}
T_1\equiv\left.\frac{\pl}{\pl\al}
\Tr \ln\{\e^{\al}<x|\e^{-t\Dvec_\m^{~\n}}|y>\}\right|_{\al=0,t=0}\com\nn
T_2\equiv -\left.\frac{\pl}{\pl\al}
\Tr \ln\{\e^{2\al}<x|\e^{-t\dvec}|y>\}\right|_{\al=0,t=0}\pr
\eea

The anomaly formula for the differential operator 
$
\Dvec_\m^{~\n}
=-g^{1/8}\{\del_\m^{~\n} \na^2+(\Ncal^\la)_\m^{~\n}\na_\la
+\Mcal_\m^{~\n}\}g^{-1/8}
$\ 
with arbitrary general covariants of $\Ncal^\la$\ and $\Mcal$ 
(taking the heat-kernel regularization
and the Fujikawa method) is derived in Appendix B and is
 given as
\bea
\label{s4}
\mbox{Anomaly}=\frac{1}{(4\pi)^2}\sqrt{g}
                \left\{\  
\mbox{Tr}[\frac{1}{6}D^2X+\half X^2
+\frac{1}{12}Y_\ab Y^\ab] \right.          \nn
\left. +4\times \frac{1}{180}
 (R_{\mn\ls}R^{\mn\ls}-R_\mn R^\mn +0\times R^2-\na^2R ) 
                \right\} \ ,\nn
X_\m^{~\n}=\Mcal_\m^{~\n}-\half (\na_\al \Ncal^\al)_\m^{~\n}
-\fourth (\Ncal^\al \Ncal_\al)_\m^{~\n}-\frac{1}{6}\del_\m^{~\n}R\ ,\nn
(Y_\ab)_\m^{~\n}=\left\{ \half (\na_\al\Ncal_\beta)_\m^{~\n}
+\fourth (\Ncal_\al \Ncal_\beta)_\m ^{~\n}-\al\change\beta\right\}
+R_{\ab\m}^{~~~~\n}\ ,\nn
(D_\al X)_\m^{~\n}=\na_\al X_\m^{~\n}+\half[\Ncal_\al,X]_\m^{~\n}\ ,\ 
(D^2X)_\m^{~\n}=\na^2X_\m^{~\n}+\half [\Ncal^\al,D_\al X]_\m^{~\n}\ .
\eea
Putting all explicit equations (\ref{s3}) into this formula, 
the vector contribution to the Weyl anomaly is finally given by
\bea
\label{s5}
T_1=\frac{1}{(4\pi)^2} \sqg\left[ 
\frac{1}{180}
 (-11 R_{\mn\ls}R^{\mn\ls}+86 R_\mn R^\mn -20 R^2 +6\na^2R) 
\right. \nn
+\{ \frac{1}{16}(P_{,\m}P^{,\m})^2+\frac{1}{12}\na^2(P_{,\m}P^{,\m})
    +\frac{5}{12}(\na^2P)^2-\frac{1}{6}P_{,\mn}P^{,\mn}   \nn
    +\frac{2}{3}R^\mn P_{,\m}P_{,\n}-\frac{1}{6}RP_{,\m}P^{,\m} 
\}                                           \nn
\left. +\{ -\frac{1}{3}\na^4P-\frac{1}{6}(P_{,\mn}+\half g_\mn \na^2P)\cdot
     P^{,\m}P^{,\n}                          
	 +\frac{1}{3}(R^\mn-\half g^\mn R)P_{,\mn}
 \}
                      \right]\pr
\eea
Terms in the first big bracket (\{\ \}) show the even power part
of $P(\p)$, while those in the second show the odd power part.
The ghost contribution to the Weyl anomaly is obtained from
(\ref{s2.9}) with $\Mcal=0$ as
\bea
\label{s6}
T_2=\frac{1}{(4\pi)^2} \sqg
\left[ 
-2\times \frac{1}{180}
 (-6\na^2R+R_{\mn\ls}R^{\mn\ls}-R_\mn R^\mn +\frac{5}{2}R^2 )
                      \right]\pr
\eea
The factor $-2$ comes from the two fields of the ghosts $c$ and
$\cbar$ and their Fermi statistics. 
There are no dilaton(P)-terms in the ghost part. 
$T_1+T_2$ gives the final result of the Weyl anomaly. It is given
by the Euler term($\sqg E$), four Weyl invariants
($I_4,I_2,I_1,I_0$)
and five trivial terms($J_3,J_{2a},J_{2b},J_1,J_0$) which are
defined in Appendix D:
\bea
\label{s7}
\mbox{Anomaly}=\frac{1}{(4\pi)^2}
\{
\frac{1}{180}(-31\sqg E+18I_0+18J_0) \nn
+\frac{1}{16}I_4+I_2+\frac{1}{3}I_{1}-\frac{1}{3}\sqg\na^4P
-\frac{1}{12}J_3+\frac{1}{12}J_{2a}+\frac{1}{3} J_{2b}
\}\com
\eea
except the term of $(-1/3)\sqg\na^4P$. In the next subsection, we show
this term, combined with the contribution from the measure
ambiguity, reduces to $(-2/15)J_1$. 
This result again confirms the general structure of the conformal
anomaly analysed in \cite{DDI76,MD77,DS93,SI98}.
A slightly generalized aspect of the above result is that
the new terms, $I_2$ and $I_{1}$, 
are conformally invariant not exactly but only up to
total derivative terms. See appendix D for details. 
The conformal anomaly for the ordinary 
photon-graviton theory (without the dilaton) is correctly reproduced for
$P(\p)=\mbox{const}$\cite{BC77,DK77,CD79}.

\subsection{Close treatment of Faddeev-Popov procedure
and the integration variables}

In the previous section, the final result has the term
$\sqg\na^4P$-term which is, by itself, not allowed by
the cohomology analysis. We miss some contribution.
First we notice one unsatisfactory treatment
of the Faddeev-Popov procedure in the present
ghost action (\ref{s2a}). It originates from the "insertion"
of the following identity in the path-integral
expression.
\bea
\label{ddddP.1}
1=\int \Del_F\del(\fg\e^{\half P(\p)}\na_\m A^\m_{\La}-K)\Dcal \La\com\nn
\Del_F=\det(\fg \e^{\half P}\na_\m\na^\m)
=\det(\invfg\e^{-\half P})\det(\sqg\e^P\na_\m\na^\m)
\com
\eea
where $K$ is a fixed constant and $\La(x)$ is the abelian
local gauge parameter (\ref{s1b}).　$\Del_F$ is the "rigorous"
F-P determinant whereas 
${\tilde \Del}_F\equiv \det(\sqg\e^P\na_\m\na^\m)$ is the
one we have taken into account by (\ref{s2a}). We must take
care of the difference factor $\det(\invfg\e^{-\half P})$, 
especially its $P$-dependent factor 
$\Del_P\equiv\det(\e^{-\half P})$ for the present interest.
( We notice the similar situation occurred in the gauge parameter
(in)dependence problem of the $\na^2 R$ anomaly-term in 4D vector model
in the background metric (no dilaton)\cite{NN88}. The careful
treatment of such a factor as above solves the problem. 
We see, from the gauge-fixing term (\ref{s2}), $\e^{\half P}$
corresponds to the "gauge parameter".
)
Furthermore we notice such factors as $\Del_P$, 
depending on the choice of the path-integral variables, can appear.
For example, in eq.(\ref{s3}), such factors, depending
on the choice of $(A_\m, \cbar)$ or $(B_\m, \bbar)$, indeed appear.
(This factor is important when the overall factor of the
path-integral get involved with the problem. 
In the case of Ref.\cite{NN88}, the
overall factor depends on the gauge parameter. 
It affects the Weyl anomaly. In the
present case, it depends on the dilaton field $P(\p)$.)
\footnote{
In the string theory, the dilaton factor $\e^{P(\p)}$
is regarded as a coupling.  In the present problem
the dilaton term
behaves like the gauge parameter.
}
As far as the dilaton is treated as the bacground field, we have
arbitrariness of such a factor. Therefore we may assume there 
exists a factor $\Del_P(\al)\equiv\det(\e^{\al P})$
, with some number $\al$, which multiplies the path-integral
we are considering. Let us evaluate its contribution to the
Weyl anomaly.

We consider the case $|P|\ll 1$ which is the region where the 
perturbative definition can be safely applied. Then the determinant
factor can be evaluated, in the heat-kernel regularization, as
\bea
\label{ddddP.2}
\ln \Del_P(\al)
=\ln\det\left\{ \e^{\al P(\p(x))}\del^{(4)}(x-y) \right\}
=\al\lim_{t\ra +0}\Tr 
\left\{ P(\p(x))<x|\e^{-t\dvec}|y> \right\} \nn
=\al\lim_{t\ra +0}\frac{1}{t}\Tr <x|\e^{-t(\dvec-P)}|y>|_{P^1}
\com
\eea
where $\dvec$ is given in (\ref{s3a}).
This trick of exponentiating the factor $P$ was used in \cite{F85}.
Let us evaluate the heat-kernel
\bea
\label{ddddP.3}
G^P(x,y;t)\equiv \Tr <x|\e^{-t(\dvec-P)}|y>\com\nn
(\frac{\pl}{\pl t}+\dvec-P)G^P(x,y;t)=0
\pr
\eea
at the order of $O(P^1)$ w.r.t. the dilaton field $P$, 
and of $O(t)$ w.r.t. the proper time $t$. 
As for the background metric, it is sufficient
to consider up to the first order of $O(h)$ w.r.t. the weak-field 
$h_\mn$ ($g_\mn=\del_\mn+h_\mn, |h_\mn|\ll 1$).
Their trace w.r.t. the space coordinates ($x=y$) 
is given by $G_1(x,x;t)$ and $G_2(x,x;t)$ appearing
in eq.(24) and eq.(25) of \cite{II96} respectively. Taking
$n=4, M=-\fourth\pl^2h+P+O(h^2), W_\mn=-h_\mn+O(h^2), 
N_\la=-\pl_\m h_{\la\m}+O(h^2)$ in the reference
we obtain the following things.
($G^P(x,x;t)=G_0(x,x;t)+G_1(x,x;t)+G_2(x,x;t)+\cdots.
$)

(i) $G_1(x,x;t)$\nl
From (24) of \cite{II96}, we obtain
\bea
\label{ddddP.4}
\frac{1}{(4\pi)^4t}\int d^4w\int^1_0dr
\frac{1}{\{(1-r)r\}^2}e^{-\frac{w^2}{4(1-r)}}\nn
\times\frac{1}{4!}\pl_\al\pl_\be\pl_\ga\pl_\del P\cdot
(\sqt)^4w^\al w^\be w^\ga w^\del
e^{-\frac{w^2}{4r}}
=\frac{1}{(4\pi)^2}\frac{t}{60}\pl^4 P
\pr 
\eea

(ii) $G_2(x,x;t)$\nl
From (25) of \cite{II96}, the relevent part of $\pl^3 h\cdot \pl P$
is given by
\bea
\label{ddddP.5}
\frac{1}{(4\pi)^6}\int d^4v\int d^4u\int^1_0dk\int_0^kdl
\frac{1}{\{(1-k)(k-l)l\}^2}e^{-\frac{v^2}{4(1-k)}}\nn
\times
\left[
\pl_\om P\sqt v^\om\e^{-\frac{(v-u)^2}{4(k-l)}}
       \left\{
\frac{1}{t}\frac{(\sqt)^3}{3!}\pl_\al\pl_\be\pl_\ga(-h_\mn)\cdot
u^\al u^\be u^\ga\frac{\pl}{\pl u^\m}\frac{\pl}{\pl u^\n}
        \right.
\right.                                        \nn
        \left.
+\frac{1}{\sqt}\frac{(\sqt)^2}{2!}\pl_\al\pl_\be(-\pl_\la h_{\m\la})
\cdot u^\al u^\be\frac{\pl}{\pl u^\m}
+\sqt\pl_\al(-\fourth\pl^2 h)\cdot u^\al
         \right\}
\e^{-\frac{u^2}{4l}}        \nn
+      \left\{
\frac{1}{t}\frac{(\sqt)^3}{3!}\pl_\al\pl_\be\pl_\ga(-h_\mn)\cdot
v^\al v^\be v^\ga\frac{\pl}{\pl v^\m}\frac{\pl}{\pl v^\n}
+\frac{1}{\sqt}\frac{(\sqt)^2}{2!}\pl_\al\pl_\be(-\pl_\la h_{\m\la})
\cdot v^\al v^\be\frac{\pl}{\pl v^\m}
        \right.         \nn
     \left.         \left.
+\sqt\pl_\al(-\fourth\pl^2 h)\cdot v^\al
                    \right\}
\e^{-\frac{(v-u)^2}{4(k-l)}}
\times\pl_\om P\cdot\sqt u^\om \e^{-\frac{u^2}{4l}}
     \right]
\nn
=\frac{t}{(4\pi)^2}     \left[
-\frac{1}{40}\pl_\m\pl^2h_{\n\n}\cdot\pl_\m P
+\frac{1}{30}\pl_\m\pl_\n\pl_\la h_{\n\la}\cdot\pl_\m P
+\mbox{other terns}     \right]
\pr 
\eea
Making use of the weak-field expansion expressions
\bea
\label{ddddP.6}
\na^4P=\pl^4P+\half\pl_\m\pl^2h\cdot\pl_\m P
-\pl^2\pl_\m h_\mn\cdot \pl_\n P
+O(h^2)\com\nn
\na_\la R\cdot P^{,\la}=\pl_\m\pl^2h\cdot\pl_\m P
-\pl_\m\pl_\n\pl_\la h_{\n\la}\cdot \pl_\m P+O(h^2)
\com 
\eea
we demand the requirement of the cohomolgy analysis, that is, 
equating the following quantity
\bea
\label{ddddP.7}
\frac{1}{(4\pi)^2}(-\third)\na^4 P+\ln \Del_P(\al)=
\frac{1}{(4\pi)^2}(-\third)\na^4 P+
\al\lim_{t\ra +0}\frac{1}{t}\Tr G^P(x,y;t)|_{P^1}\nn
=\frac{1}{(4\pi)^2}(  (\frac{\al}{60}-\third)\pl^4 P
-(\frac{\al}{40}+\frac{1}{6})\pl_\m\pl^2 h\cdot\pl_\m P
+\frac{\al}{30}\pl_\m\pl_\n\pl_\la h_{\n\la}\cdot\pl_\m P\nn
+\mbox{other terms}
                              )
\com 
\eea
with the trivial term
\bea
\label{ddddP.8}
\be\times J_1=\be\sqg (3\na^4 P-\na^\la(RP_{,\la}))\nn
=\be\left\{
3\pl^4P+\half\pl_\m\pl^2h\cdot P_{,\m}
+\pl_\m\pl_\n\pl_\la h_{\n\la}\cdot P_{,\m}
\right\}+\mbox{other terms}
\com 
\eea
we find
\bea
\label{ddddP.9}
\al=-4\com\q \be=\frac{1}{(4\pi)^2}\times (-\frac{2}{15})
\pr 
\eea
The consistency with the cohomology determines the
dilaton-depending
overall factor of the path-integral.
\section{Trace anomaly for 4D dilaton coupled spinor}
The dilaton can be introduced into the (Dirac) spinor-gravity
system keeping the Weyl invariance as follows:
\bea
\label{sp.1}
S[\psi,\psibar;g,\p]=\half\intfx \sqg \e^{P(\p)}\{\psibar i\overnab \psi\}\ ,\ 
 {\bar \psi}\overnab \psi = {\bar\psi} ( \gamma^\mu \nabla_\mu \psi ) -
( {\bar \psi}  {\overleftarrow\nabla}_\mu \gamma^\mu) \psi\ ,          \nn
\nabslash = \gamma^\mu \nabla_\mu\com\q
 \nabla_\mu \psi = (\partial_\mu + \omega_\mu ) \psi\com\q
 {\bar\psi} {\overleftarrow\nabla}_\mu = {\bar\psi} 
({\overleftarrow\partial}_\mu - \omega_\mu ),                         \nn 
 \omega_\mu{}^a{}_b = ( \Gamma^\lambda{}_{\mu\nu} e_\lambda{}^a 
	- \partial_\mu e_\nu{}^a ) e^\nu{}_b\com\q
 \omega_\mu = \half \sigma^{ab} \omega_{\mu ab}\pr  
\eea
It is invariant under the Weyl transformation:
\bea
\label{sp.2}
\psibar'=\e^{-\frac{3}{2}\al}\psibar\com\q
\psi'=\e^{-\frac{3}{2}\al}\psi\com\q
{\e_\m^{~a}}'=\e^\al\e_\m^{~a}\com\q \p'=\p\pr
\eea
We can absorb all $P(\p)$ dependence by the rescaling of fields:
\bea
\label{sp.3}
\e^{\half P(\p)}\psi=\chi\com\q
\e^{\half P(\p)}\psibar=\chibar\com\nn
S=\half\intfx \sqg \chibar i\overnab \chi
=\half\intfx {\tilde \chibar}\fg i\overnab \invfg{\tilde \chi}\com
\eea
where ${\tilde \chibar}=\fg\chibar$ and ${\tilde \chi}=\fg\chi$.
Therefore the dilaton does {\it not} change
the conformal anomaly in the spinor-gravity theory
as far as we take the appropriate measure:
$\Dcal{\tilde \chibar}\Dcal{\tilde \chi}=
\Dcal(\fg\e^{P/2}\psi)\Dcal(\fg\e^{P/2}\psibar)$.

\section{Induced Action for Dilatonic Wess-Zumino model}
In the present section we apply the previous results to
the dilatonic Wess-Zumino (WZ) model and
discuss the conformal anomaly and the anomaly induced action. 
The model may be constructed following the book by
Wess and Bagger\cite{WB92}. We suppose that the model 
is coupled with the external gravitational multiplet
and the external dilaton
(chiral) multiplet $\Psi$. Then using the notation of above book
we may write the conformally invariant action:
\bea
\label{wz.1}
S=\int d^2\Th~ 2\Ecal[-\frac{1}{16}({\bar \Dcal}{\bar \Dcal}+8R)
\{ \Phi^\dag(\e^{P(\Psi)}\Phi)+\e^{P(\Psi^\dag)}\Phi^\dag\Phi \}
                     ]\ ,
\eea
where $\Phi=A+\sqrt{2}\Th\chi+\dots$ and
$\Psi=\p+\cdots$.
Choosing only the bosonic background for $\Psi$ and
eliminating the auxiliary fields leads to the following
component form of WZ action. 
\bea
\label{wz.2}
L=e~ \e^{P(\p)}\{ -A\na^2 A^*-\frac{1}{6}RAA^*
                 -\frac{i}{2}(\chi\si^\m \Dcal_\m {\bar \chi}
+{\bar \chi}\si^\m\Dcal_\m\chi) 
                \}\ ,
\eea
where we take the choice $\p^*=\p$ for the dilaton. $A$
is the complex scalar and $\chi$ is the Majorana spinor.

The conformal anomaly for the system may be found as follows
\bea
\label{wz.3}
& T=\frac{1}{(4\pi)^2}
\{
b I_0+b'\sqg E+b'' J_0
+a I_4+a' J_{2a}
\}\com                                          &   \nn
& a=\frac{1\times 2}{32}=\frac{1}{16}\ ,\ 
a'=\frac{1\times 2}{24}=\frac{1}{12}\ ,\      
b=\frac{1}{120}(2+6\times\half)=\frac{1}{24}\ ,  & \nn 
& b'=-\frac{1}{360}(2+11\times\half)=-\frac{1}{48}\ ,\  
b''=-\frac{1}{180}(2+6\times\half)=-\frac{1}{36}\ . &
\eea
$b''$ and $a'$ are ambiguous as they may be changed by 
some finite counterterms in the gravitational action.

The nonlocal effective action may be found by integrating
the Weyl-transformed anomaly equation:
\footnote{
See \cite{RR84,FT84,BOS85} for the purely gravitational case
and \cite{NO6143} for the dilaton-gravitational case.
}
\bea
\label{wz.4}
T(\e^{2\si}g_\mn)=\half\frac{\del}{\del\si}W(\e^{2\si}g_\mn).
\eea
The final expression for the anomaly induced effective action
$W$ is given as 
\bea
\label{wz.5}
& W(g_\mn)=\frac{1}{\fpisq}\int d^4x d^4y 
\{ bI_0+aI_4+\frac{b'}{2}(\sqg E+\frac{2}{3}J_0) \}_x
G(x,y)\{\fourth(\sqg E+\frac{2}{3}J_0)\}_y   & \nn
& +\frac{1}{\fpisq}\intfx\{ 
\frac{1}{12}(b''-\frac{2}{3}b')K_0
+\frac{a'}{6}K_{2a}         \}\ ,  &
\eea
where 
$W$ is in a covariant but non-local form via
\bea
\label{wz.6}
(\sqg\Delta)_xG(x,y)=\del^4(x-y)\ ,\ 
\Delta=(\na^2)^2-2R^\mn\na_\m\na_\n+\frac{2}{3}R\na^2
-\third(\na^\m R)\na_\m\ .
\eea
Here $\Delta$ is the conformally invariant 4-th order differential
operator (Appendix D). 
For the case $a=a'=0$, the result (\ref{wz.5}) coincides with 
Ref.\cite{RR84}.
Note that it is the effective action (\ref{wz.5}) that
may serve as a starting point for constructing IR sector
of the dilatonic gravity in the analogy with the purely
gravity case of \cite{AM92,SO92}.

From another point using eq.(\ref{wz.5}) one can study the
backreaction problem for the dilatonic WZ model. Here new possibilities
are expected to appear in constructing the inflationary universe
induced by the model. We hope to return to this question in more detail
in future.

\section{Conclusion}
We have shown  how the dilaton affects the conformal anomaly
in various gravity-matter theories. The dilaton has the same status
as the gravitational field. 
We summarize the results of this paper by
writing the anomaly $T_S,T_V$ and $T_F$ for the scalar, the vector
and the Dirac fermion respectively:
\bea
\label{conc.0}
& T_S=\frac{1}{(4\pi)^2}
\{
\frac{1}{360}(-\sqg E+3I_0)
+\frac{1}{32} I_4+\mbox{Trivial Terms}
\}\com  &                                        \nn
& T_V=\frac{1}{(4\pi)^2}
\{
\frac{1}{180}(-31\sqg E+18I_0)
+\frac{1}{16} I_4+I_2+\frac{1}{3}I_{1}+\mbox{Trivial Terms}
\}\com   &                                       \nn
& T_F=\frac{1}{(4\pi)^2}
\{
\frac{1}{180}(-\frac{11}{2}\sqg E+9I_0)
+\mbox{Trivial Terms}
\}\pr  &                                        
\eea
For the system with 
$N_S$ scalars, $N_F$ Dirac fermions and $N_V$
vectors, the conformal anomaly is given by 
\bea
\label{conc.1}
& \mbox{Anomaly}=\frac{1}{(4\pi)^2}
\{
b_E\sqg E+b_0 I_0
+b_4 I_4+b_2 I_2+b_{1} I_{1}+\mbox{Trivial Terms}
\}\ ,                                         & \nn
& b_E=\frac{1}{360}(-N_S-62N_V-11N_F)<0\ ,\ 
b_0=\frac{1}{120}(N_S+12N_V+6N_F)>0\ ,         & \nn
& b_4=\frac{1}{32}(N_S+2N_V+0\times N_F)>0\ ,\ 
b_2=0\times N_S+N_V+0\times N_F>0\ ,           & \nn 
& b_{1}=0\times N_S+\frac{1}{3}N_V+0\times N_F>0\ 
 . &
\eea
This is the dilaton generalization of a result in Ref.\cite{MD77}.
"Trivial terms" depend on the choice of regularization, 
measure and  finite counterterms\cite{SI97}. Their meaning is not
very clear ( at least, in the perturbative approach).
\footnote{
The gauge (in)dependence of the "trivial terms" 
was analysed in Ref.\cite{RE84} and
the gauge independence was explicitly shown in Ref.\cite{NN88}.
} 
On the other hand, the coefficients $b$'s are clearly obtained
 and  depend only on the content of
the matter fields. They remind us of the critical indices
in condensed matter physics.

In Sec.5 we have presented the anomaly induced action for the dilatonic
Wess-Zumino model by integrating the conformal anomaly equation
w.r.t. the Weyl mode. It is known that some anomaly equations 
like those of QED or $\la\phi^4$ theory on the curved space (with 
account of matter interaction) can not
be integrated because of the presence of $R^2$-term in the anomaly
\cite{RR84}. Recently, in Ref.\cite{NO98} it is noted that
the anomaly for the {\it free} photon on the dilatonic 
curved space (\ref{s7}), 
cannot be integrated because of the presence of the new type (generalized) 
conformal invariants $I_2,I_{1a}$ and $I_{1b}$. In other words,
 a covariant anomaly induced effective action cannot be constructed 
for a dilaton coupled vector.
It indicates some speciality of the
gauge theory in comparison  
with the scalar and fermion fields. It could be 
possible to construct only non-covariant anomaly-induced 
effective action in the dilaton-coupled vector case. Probably, such 
an action 
still could lead to a correct conformal anomaly like in the case with 
Chern-Simons invariants.
It is also interesting to note that our results maybe well applied 
to SUSY dilaton coupled theories like Wess-Zumino model (for two-dimensional 
case, see \cite{O}).
 
If string theory or other alternative theory (like M-theory)
takes over
local field theory, the latter formalism should show some
theoretical breakdown at some scale such as the Planck mass.
At present some pathological phenomena, such as 
the entropy loss problem in the black hole geometry could 
look like such a breakdown. We note here that 
the conformal anomaly is considered to be a key factor in the
Hawking radiation phenomenon. 
In the transitive period from local field theory
to string field theory,  "intermediate" fields such
as the dilaton are thought to
play an important role in understanding
 new concepts involved in the Planck mass physics. 
The Weyl anomaly occurs quite generally for most 
ordinary conformally invariant field theories.
It shows that
the quantization of fields necessarily breaks the
scale invariance. 
It is always related with the ultraviolet regularization.
On the other hand
it is frequently suggested that  string
theory could predict some minimal unit of length in space(-time). 
The effect of such a unit length on its field theory limit
could be related to the conformal anomaly phenomenon.
We hope the present results are helpful in studying of such problems.

\newpage
{\large Appendix A. Anomaly Formula for Scalar Theories
on Curved Space}

\vs{0.5}
In this appendix we derive an anomaly formula for the general
differential operator
\bea
\label{a1}
\Dvec=\fg (-\na^2\cdot{\bf 1}-\Mcal)\invfg\com
\eea
which frequently appears in the scalar theories on curved
space. ${\bf 1}$ is the unit matrix with the size of the field
indices:\ $({\bf 1})_{ij}=\del_{ij}; i,j=1,2,\cdots,N_s$.
We proceed in three steps.
\footnote{
This approach was taken in the derivation of the counterterm 
formula\cite{IO82}.
}

\vs{0.3}
i)\ Flat space\nl
For the case of the flat space :\ $g_\mn=\del_\mn$ the operator
reduces to $-\pl^2-\Mcal$ and the anomaly is known 
\cite{II96,DAMTP9687} as
\bea
\label{a2}
\mbox{Anomaly}=\frac{1}{\fpisq}\Tr 
(\sixth\pl^2\Mcal+\half \Mcal^2)\pr
\eea
where "$\Tr$" means "trace" over the field suffixes $i$:\ 
$\Tr \Mcal=\sum_{i=1}^{N_s}\Mcal_{ii}$. 

\vs{0.3}
ii)\ General curved space\nl
 From the general covariance the most general form which
reduces to (\ref{a2}) in the flat space can be expressed as
\bea
\label{a3}
\mbox{Anomaly}=\frac{\sqg}{\fpisq}\Tr \{\sixth\na^2 X+\half X^2\nn
+(a_1R_{\mn\ls}R^{\mn\ls}+a_2R_\mn R^\mn+a_3R^2+a_4\na^2 R)
\cdot{\bf 1} \}                                        \com\nn
X=\Mcal+c_1R\com
\eea
where $a_1\cdots a_4$ and $c_1$ are some constants to be determined.

\vs{0.3}
iii)\ Determination of constants\nl
First we consider the case $\Mcal=\frac{1}{6}R$ which corresponds to
the case:$f(\p)=1,\xi=-\sixth$ in Sec.2. The above
result (\ref{a3}) says
\bea
\label{a4}
N_s=1\com\q\Mcal=\sixth R\com\q X=(\sixth+c_1)R\com\nn
\mbox{Anomaly}=\frac{\sqg}{\fpisq} \left(
\{\sixth(\sixth+c_1)+a_4\}\na^2 R\right.\nn
\left.+a_1R_{\mn\ls}R^{\mn\ls}+a_2R_\mn R^\mn
+\{\half(\sixth+c_1)^2+a_3\}R^2
                                  \right)\pr
\eea
Comparing this with the well-established result (say, eq.(40) of
\cite{II96})
\bea
\label{a5}
\mbox{Anomaly}=\frac{\sqg}{\fpisq} \frac{1}{180}
(-\na^2 R+R_{\mn\ls}R^{\mn\ls}-R_\mn R^\mn )\pr
\eea
we obtain the first set of constraints on the constants.
\bea
\label{a6}
a_1=\frac{1}{180}\com\q a_2=\frac{-1}{180}\com\nn
\sixth(\sixth+c_1)+a_4=\frac{-1}{180}\com\q
\half (\sixth+c_1)^2+a_3=0       \pr
\eea

Next we consider the case of the weak gravity 
($g_\mn=\del_\mn+h_\mn,\ |h_\mn|\ll 1$)
for the general
operator (\ref{a1}) in order to fix the value of $c_1$.
The operator is expanded as
\bea
\label{a7}
& \Dvec=-{\bf 1}\pl^2-\Mcal
+{\bf 1}(h_\mn\pl_\m\pl_\n+\pl_\m h_{\m\la}\pl_\la+\fourth\pl^2h)
+O(h^2) &             \nn
& \equiv  -{\bf 1}\pl^2-W_\mn\pl_\m\pl_\n-N_\la\pl_\la-M\pr & \nn  
& W_\mn=-h_\mn+O(h^2)\ ,\ N_\la=-\pl_\m h_{\m\la}+O(h^2)\ ,\ 
M=\Mcal-\fourth\pl^2h+O(h^2)\ . &
\eea
Now we focus on the term $(c_1/\fpisq)\sqg\Mcal R=
(c_1/\fpisq)\Mcal(\pl^2h-\pl_\m\pl_\n h_\mn)+O(h^2)$ in (\ref{a3}).
The conformal anomaly from (\ref{a7}) is independently
obtained by the result of \cite{II96}
((38),Table I) taking $W_\mn,N_\la,M$ in the reference as given
above:
\bea
\label{a8}
\frac{1}{\fpisq}\Tr (-\frac{1}{12}M\pl^2W_{\m\m}
+\frac{1}{3}M\pl_\m\pl_\n W_\mn-\half M\pl_\m N_\m
+\half M^2) \nn
\sim \frac{-1}{\fpisq}\Tr\Mcal(\sixth\pl^2 h-
\sixth\pl_\m\pl_\n h_\mn+O(h^2))
=\frac{1}{\fpisq}\sqg (\frac{-1}{6})\Tr\Mcal R  \pr
\eea
This shows
\bea
\label{a9}
c_1=-\sixth\pr
\eea
(\ref{a6}) and (\ref{a9}) fix all constants. Finally we obtain
the anomaly formula for (\ref{a1}) as
\bea
\label{a10}
\mbox{Anomaly}=\frac{\sqg}{\fpisq}\Tr \{ \sixth\na^2 X+\half X^2\nn
+\frac{1}{180}(R_{\mn\ls}R^{\mn\ls}-R_\mn R^\mn
+0\cdot R^2-\na^2 R)\cdot {\bf 1}
                                       \}\com\nn
X=\Mcal-\sixth R\cdot{\bf 1}\pr
\eea
As noted in the text, this result ( and (\ref{s4}) to be derived
in the next appendix ) is the same as the counterterm formula
(say of ref. \cite{BV81,BV85,BOS92}) where dimensional regularization was taken.

\vspace{2cm}
{\large Appendix B. Anomaly Formula for Vector Theories
on Curved Space}

\vs{0.5}
In this appendix we derive the anomaly formula for the operator
\bea
\label{b1}
\Dvec_\m^{~\n}
=-g^{1/8}\{ \del_\m^{~\n} \na^2+(\Ncal^\la)_\m^{~\n}\na_\la
+\Mcal_\m^{~\n} \}g^{-1/8}\com
\eea 
which generally appears  for the vector fields on the curved
background. $\Ncal^\la$\ and $\Mcal$\ are general covariants. We proceed
in the similar way as in App.A.

\vs{0.3}
i)Flat Space\nl
For the case of flat space $g_\mn=\del_\mn$, the formula appears 
in \cite{II96,DAMTP9687}:
\bea
\label{b2}
\Dvec_\m^{~\n}
=-\del_\m^{~\n} \pl^2-(\Ncal^\la)_\m^{~\n}\pl_\la-\Mcal_\m^{~\n}\com   \nn
X_\m^{~\n}=\Mcal_\m^{~\n}-\half(\pl_\la\Ncal^\la)_\m^{~\n}
-\fourth(\Ncal_\la\Ncal^\la)_\m^{~\n}\com\nn
(Y^\mn)_\al^{~\be}=\half(\pl^\m\Ncal^\n-\pl^\n\Ncal^\m)_\al^{~\be}
+\fourth[\Ncal^\m,\Ncal^\n]_\al^{~\be}\com\nn
D^\m X_\al^{~\be}=\pl^\m X_\al^{~\be}+\half [\Ncal^\m,X]_\al^{~\be}\com\nn
\mbox{Anomaly}_{\mbox{flat}}=\frac{1}{\fpisq}\Tr[\sixth D^2X+\half X^2
+\frac{1}{12}Y_\mn Y^\mn ]\com
\eea 
where the upper and lower indices do not have meaning in present
case ($g_\mn=g^\mn=\del_\m^\n$ are all the Kronecker's delta), but
only do have a notational continuity with the next general case.

\vs{0.3}
ii)\ Curved Space \nl
The anomaly formula can be first expressed as the most general form
which reduces to (\ref{b2}) at the flat space limit:
\bea
\label{b3}
X_\m^{~\n}=\Mcal_\m^{~\n}-\half(\na_\la\Ncal^\la)_\m^{~\n}
-\fourth(\Ncal_\la\Ncal^\la)_\m^{~\n}+c_1\del_\m^\n R\com\nn
(Y^\mn)_\al^{~\be}=\half(\na^\m\Ncal^\n-\na^\n\Ncal^\m)_\al^{~\be}
+\fourth[\Ncal^\m,\Ncal^\n]_\al^{~\be}+c_2R^{\mn~\be}_{~~\al}\com\nn
D^\m X_\al^{~\be}=\na^\m X_\al^{~\be}+\half [\Ncal^\m,X]_\al^{~\be}\com\nn
\mbox{Anomaly}=\frac{\sqg}{\fpisq}\Tr \{\sixth D^2 X+\half X^2
+\frac{1}{12}Y_\mn Y^\mn\nn
+(a_1R_{\mn\ls}R^{\mn\ls}+a_2R_\mn R^\mn+a_3R^2+a_4\na^2 R)
\cdot{\bf 1} \}                                        \com
\eea
where $c_1,c_2,a_{1-4}$ are some constants to be determined.

\vs{0.3}
iii)\ Determination of constants

\vs{0.3}
iiia)\ First we consider a special case of 
$(\Ncal^\la)_\m^{~\n}=0,\Mcal_\m^{~\n}=R_\m^{~\n}$,
which is a special case of $f(\p)=1$ in (\ref{s1}) and is
the ordinary photon-gravity theory. Comparing the result from (\ref{b3})
\bea
\label{b4}
X_\m^{~\n}=R_\m^{~\n}+c_1\del_\m^\n R\com\ \ \ 
(Y^\mn)_\al^{~\be}=c_2R^{\mn~\be}_{~~\al}\com\nn
\mbox{Anomaly}=\frac{\sqg}{\fpisq} \{\sixth\na^2 (R+4c_1R)
+\half (R_\m^{~\n}+c_1\del_\m^\n R)^2   \nn
+\frac{(c_2)^2}{12}R^{\mn~\be}_{~~\al} R_{\mn\be}^{~~~\al}
+4a_1R_{\mn\ls}R^{\mn\ls}+4a_2R_\mn R^\mn+4a_3R^2+4a_4\na^2 R \}\com  
\eea
with the well-established result (say, eq.(2.14) of \cite{RE84})
\bea
\label{b5}
\mbox{Anomaly}=\frac{\sqg}{\fpisq} \{
-\frac{11}{180}R_{\mn\ls}R^{\mn\ls}+\frac{43}{90}R_\mn R^\mn
-\frac{1}{9}R^2+\frac{1}{30}\na^2 R \}\com                                 
      \com
\eea
we obtain
\bea
\label{b6}
-\frac{1}{12}(c_2)^2+4a_1=-\frac{11}{180}\com\q
\half+4a_2=\frac{43}{90}\com\nn
\half (2c_1+4{c_1}^2)+4a_3=-\frac{1}{9}\com\q
\frac{1+4c_1}{6}+4a_4=\frac{1}{30}\pr
\eea

We give here a list of the weak-field
expansion of the operator (\ref{b1}) for the use in iiib) and iiic).
\bea
\label{b12}
g_\mn=\del_\mn+h_\mn\com\q |h_\mn|\ll 1\ ,\nn
\Dvec_\al^{~\be}=-\del_\al^\be\pl^2-(W_\mn)_\al^{~\be}\pl_\m\pl_\n
-(N_\m)_\al^{~\be}\pl_\m-M_\al^{~\be}\ ,\ 
(W_\mn)_\al^{~\be}=-\del_\al^{~\be} h_\mn+O(h^2)\ ,         \nn 
(N_\m)_\al^{~\be}=(\Ncal^\m)_\al^{~\be}
-\del_\al^\be(\pl_\la h_{\la\m}-\fourth\pl_\m h)   
-(\pl_\m h_\ab+\pl_\al h_{\m\be}-\pl_\be h_{\m\al})+O(h^2)\ ,  \nn 
M_\al^{~\be}=\Mcal_\al^{~\be}-\frac{1}{8}\del_\al^\be\pl^2h
-\half\pl_\m(\pl_\m h_\ab+\pl_\al h_{\m\be}-\pl_\be h_{\m\al})  \nn
-\frac{1}{8}(\Ncal^\la)_\al^{~\be}\pl_\la h
-\half(\Ncal^\la)_\al^{~\tau}(\pl_\la h_{\be\tau}+\pl_\tau h_{\la\be}
-\pl_\be h_{\la\tau})+O(h^2)\ .
\eea

\vs{0.3}
iiib)\ Second we consider the case $(\Ncal^\la)_\m^{~\n}=0$
with a general $\Mcal_\m^{~\n}$:\ 
$\Dvec_\m^{~\n}=-g^{1/8}\{\del_\m^{~\n} \na^2+\Mcal_\m^{~\n}\}g^{-1/8}$.
We focus on the $R\Tr \Mcal$ term. The formula (\ref{b3}) gives
\bea
\label{b7}
\frac{c_1}{\fpisq}R\Tr \Mcal\pr
\eea
On the other hand, the anomaly can be independently obtained by
the weak-field expansion (\ref{b12}) and the
anomaly formula
eq.(38) and TABLE I of ref.\cite{II96}:\ 
\bea
\label{b9}
\frac{1}{(4\pi)^2}\Tr (-\frac{1}{12}M\cdot\pl^2W_{\m\m}
+\frac{1}{3} M\cdot\pl_\m\pl_\n W_\mn-\half M\cdot\pl_\m N_\m
+\half M^2)  \nn
\sim\frac{1}{(4\pi)^2}\times \frac{1}{6}
(-\pl^2h+\pl_\m\pl_\n h_\mn)\times\Tr\Mcal
= \frac{1}{(4\pi)^2}\times \frac{-1}{6}(R+O(h^2))\Tr\Mcal
\ ,
\eea
and we obtain
\bea
\label{b10}
c_1=-\sixth\pr
\eea

\vs{0.3}
iiic)\ With the purpose of determining  $c_2$, we consider
the $\Mcal=0$ case with a general $\Ncal^\la$:\ 
$\Dvec_\al^{~\be}
=-g^{1/8}\{\del_\al^{~\be} \na^2+(\Ncal^\la)_\al^{~\be}\na_\la\}g^{-1/8}$.
We focus on the $ (\Ncal_\m\Ncal_\n)_\ab R^{\mn\be\al} $
term in the anomaly. From (\ref{b3})
we have 
\bea
\label{b11}
+\frac{c_2}{12}(\Ncal_\m\Ncal_\n)_\ab R^{\mn\be\al}\pr
\eea
We can derive the above result independently in the weak gravity
expansion (\ref{b12}).
Eq.(43) of \cite{DAMTP9687} gives, as the corresponding 
$\Ncal\Ncal\pl\pl h$
part,
\bea
\label{b13}
& \frac{1}{(4\pi)^2}\frac{1}{12}\Tr Y_\mn Y_\mn
\sim\frac{1}{(4\pi)^2}\times (-\frac{1}{12})\times
\half(\pl_\m \pl_\al h_{\n\be}-\pl_\m\pl_\be h_{\n\al}-\m\change\n)
(\Ncal^\m\Ncal^\n)_\be^{~\al}  & \nn
& \sim\frac{1}{(4\pi)^2}\times
(+\frac{1}{12})R_{\mn\al}^{~~~\be}(\Ncal^\m\Ncal^\n)_\be^{~\al}\com &
\eea
where $Y_\mn\equiv \pl_\m A_\n-\pl_\n A_\m+[A_\m,A_\n],\ 
A_\m\equiv\half (\del_\mn+Z_\mn)N_\n$ and
$(\del_\mn+W_\mn)(\del_{\n\la}+Z_{\n\la})=\del_{\m\la}\cdot {\bf 1}$.
Therefore we obtain
\bea
\label{b14}
c_2=+1\pr
\eea
Finally we obtain
\bea
\label{b15}
a_1=+\frac{1}{180}\ ,\ a_2=-\frac{1}{180}\ ,\ a_3=0\ ,\ 
a_4=-\frac{1}{180}\ ,\ 
c_1=-\frac{1}{6}\ ,\ c_2=+1\com
\eea
and the general formula is given by (\ref{s4}).

\vspace{2cm}
{\large Appendix C. Weyl Transformation }

\vs{0.5}
We list here some formulae of Weyl transformation for
various fields. Only in this appendix the space(-time)
is n-dimensional.

\vs{0.3}
i)\ Definition\nl
\bea
\label{c1}
R^\la_{~\m\n\si}=\pl_\n\Ga^\la_{\m\si}+\Ga^\la_{\tau\n}\Ga^\tau_{\m\si}-
\n\change\si\ ,\ 
\Ga^\la_\mn=\half g^\ls (\pl_\m g_{\si\n}+\pl_\n g_{\si\m}-\pl_\si g_\mn)\ ,\nn
R_\mn =R^\la_{~\mn\la}\ ,\  R=g^\mn R_\mn\ ,\  g=+\mbox{det}g_\mn\ ,\ 
\nabla_\mu A_\nu = \partial_\mu A_\nu - \Gamma^\lambda_{\mu\nu} 
A_\lambda\ .
\eea

\vs{0.3}
ii)\ Weyl transformation of gravitational fields\nl
\bea
\label{c2}
g'_\mn=\e^{2\al(x)}g_\mn\com\q \sqrt{g'}=\e^{n\al}\sqg\com\q
{\Ga^\la_\mn}'-\Ga^\la_\mn=\del^\la_\n\al_{,\m}+\del^\la_\m\al_{,\n}
-g_\mn\al^{,\la}\ ,\ 
\eea
where $\al_{,\m}=\na_\m\al(=\pl_\m\al),\al^{,\la}=g^{\ls}\al_{,\si}$.
This is the finite transformation, not the infinitesimal one:
\bea
\label{c3}
{R^\la_{~\mn\si}}'-R^\la_{~\mn\si}=      
\del^\la_\si
( \al_{,\mn}-\al_{,\m}\al_{,\n}+g_{\mn}\al_{,\tau}\al^{,\tau})
+g_\mn ( \al^{,\la}_{~,\si}-\al^{,\la}\al_{,\si})
-(\n\change\si)\ ,                                 \nn
{R_\mn}'-R_\mn=  g_\mn\na^2\al+     
(n-2) (\al_{,\mn}-\al_{,\m}\al_{,\n}+g_\mn\al_{,\la}\al^{,\la})\ ,\nn
R'=\e^{-2\al}\{ R+2(n-1)\na^2\al+(n-1)(n-2)\al_{,\m}\al^{,\m} \}\ ,
\eea
where $\al_{,\mn}=\na_\n\na_\m\al$.

\vs{0.3}
iii)\ Weyl tensor\nl
The Weyl tensors defined below are invariant 
under the Weyl transformation:
\bea
\label{c4}
C^\la_{~\mn\si}=R^\la_{~\mn\si}+\frac{1}{n-2}(\del^\la_\n R_{\m\si}
+g_{\m\si}R^\la_{~\n}-\n\change\si)\nn
+\frac{1}{(n-2)(n-1)}(\del^\la_\si g_\mn-\n\change\si)R\com\nn
{C^\la_{~\mn\si}}'=C^\la_{~\mn\si}\com\q
C_{\la\mn\si}=g_{\la\tau}C^\tau_{~\mn\si}\com\q
{C_{\la\mn\si}}'=\e^{2\al}C_{\la\mn\si}\pr
\eea

\vs{0.3}
iv)\ Scalar
\bea
\label{c5}
\vp'=\e^{-\frac{n-2}{2}\al(x)}\vp\com\q
(\na_\m\vp)'=\e^{-\frac{n-2}{2}\al(x)}(\na_\m\vp
-\frac{n-2}{2}\al_{,\m}\vp)\com                     \nn
(\na^2\vp)'=\e^{-\frac{n+2}{2}\al(x)}\{
\na^2\vp-\frac{n-2}{2}\vp\na^2\al
-\frac{(n-2)^2}{4}\al_{,\m}\al^{,\m}\vp\}\pr
\eea
The well-known conformal coupling is recognized from the relation:
\bea
\label{c6}
\left[\sqg(\vp\na^2\vp+\frac{n-2}{4(n-1)}R\vp^2)\right]'
=\sqg(\vp\na^2\vp+\frac{n-2}{4(n-1)}R\vp^2)\pr
\eea

\vs{0.3}
v)\ Dilaton
\bea
\label{c7}
\p'=\p\ ,\ 
(\na_\m\p)'=\na_\m\p(=\pl_\m\p)\ ,\ 
(\na^2\p)'=\e^{-2\al}\{\na^2\p+(n-2)\al^{,\la}\p_{,\la}\}\ .
\eea
We note the dilaton and the scalar transform in the same way
only in the 2D space:
\bea
\label{c8}
(\na_\m\p\cdot\na^\m\p)'=\e^{-2\al}\na_\m\p\cdot\na^\m\p\com\nn
(\na_\m\na_\n\p)'=\na_\m\na_\n\p-(\al_{,\m}\p_{,\n}
+\al_{,\n}\p_{,\m}-g_\mn\al^{,\la}\p_{,\la})\pr
\eea
 Here we see $\sqg(\na_\m\p\cdot\na^\m\p)^{n/2}$
is conformal invariant in n-dim space. In the text the general
form of the dilaton coupling $\e^{P(\p)}$ is used instead of
$\e^{\mbox{const}\times\p}$. 
We may substitute
$P(\p)$ for $\p$ in the above formulae  
(e.g.\ $
{P_{,\mn}}'=P_{,\mn}-(\al_{,\m}P_{,\n}
+\al_{,\n}P_{,\m}-g_\mn\al^{,\la}P_{,\la})$).

\vspace{2cm}
{\large Appendix D. Weyl Invariants and Trivial Terms }

\vs{0.5}
We consider 4D space in this appendix.
In the present model we have the following conformal invariants
made only of the metric and the dilaton field:
\bea
\label{inv1}
& I_0\equiv\sqg C_{\mn\ls}C^{\mn\ls}\com\q
I_4\equiv\sqg(P_{,\m}P^{,\m})^2\com                      & \nn
& I_2\equiv\sqg(R^\mn P_{,\m}P_{,\n}-\frac{1}{6}RP_{,\m}P^{,\m}
+\frac{3}{4}(\na^2P)^2-\half P^{,\mn}P_{,\mn})\com       & \nn
& I_{1}\equiv\sqg(R^\mn-\half g^\mn R)P_{,\mn}
=\sqg\na_\n\{ (R^\mn-\half g^\mn R)P_{,\m} \} \com       & 
\eea
where the numbers in the lower suffixes show the numbers 
of powers $P$. They transform under the finite Weyl transformation as
\bea
\label{inv2}
& {I_0}'=I_0\com\q {I_4}'=I_4\com & \nn
& {I_2}'-I_2=2\sqg\na^\m(\al^{,\n}P_{,\m}P_{,\n})\com & \nn
& {I_{1}}'-I_{1}=\sqg\na^\n
\{ 2(P_{,\mn}\al^{,\m}-\na^2P\cdot\na_\n\al)
  -(2\al^{,\m}\al_{,\n}P_{,\m}+\al_{,\m}\al^{,\m}P_{,\n}) \}\pr & 
\eea
Two terms $I_2$ and $I_{1}$, 
which are conformally invariant up to total derivative terms, 
may be called {\it generalized} conformal invariants. They lead to strange
consequences, 
in comparison with standard conformal invariants, in construction 
of the anomaly induced effective action. As the quantities with the same
properties,  
we note the Euler term, $\sqg E$, defined in (\ref{s2.12}) and a trivial term,
$J_0$, defined below.
\bea
\label{inv3}
& (\sqg E)'-\sqg E=4\sqg\na^\m \{ R\al_{,\m}-2R_\m^{~\n}\al_{,\n}  
 -\na_\m(\al_{,\n}\al^{,\n})+2\al_{,\m}\na^2\al
                          +2\al_{,\n}\al^{,\n}\al_{,\m}\}\ ,       & \nn
& \frac{2}{3}({J_0}'-J_0)=4\sqg\na^\m\{ \na_\m(\na^2\al+\al_{,\n}\al^{,\n}) 
 -\al_{,\m}(\third R+2\na^2\al+2\al_{,\n}\al^{,\n})\}\ ,  &  \nn
& (\sqg E+\frac{2}{3}J_0)'-(\sqg E+\frac{2}{3}J_0)=4\sqg \Del\al  \ , &
\eea
where $\sqg\Del$ is the conformally invariant 4-th order differential operator
defined in eq. (\ref{wz.6}) of the text.

Among the Weyl anomaly terms there are those terms which can be obtained by the
infinitesimal Weyl transformation
of local (counter)terms composed only of the background
fields $(g_\mn,\p)$. They are called "trivial" (anomaly) terms
\cite{DS93}. ( In the mathematical (cohomology) terminology,
it is called {\it coboundary} \cite{BPB}).
To obtain them, it is sufficient to consider
the infinitesimal transformation($\al(x)\ll 1$) in
the appendix C: 
\bea
\label{tri1}
& J_0\equiv\sqg\na^2R\ ,\ 
K_0\equiv\sqg R^2\ ,\ 
g_\mn\frac{\del}{\del g_\mn (x)}\intfx K_0=6 J_0\ , & \nn
& J_{2a}\equiv\sqg\na^2(P_{,\m}P^{,\m})\ ,\ 
K_{2a}\equiv\sqg R P_{,\m}P^{,\m}\ ,\ 
g_\mn\frac{\del}{\del g_\mn (x)}\intfx K_{2a}=+3 J_{2a}\ , &\nn
& J_{2b}\equiv\sqg(-R^\mn P_{,\m}P_{,\n}+P_{,\mn}P^{,\mn}-(\na^2P)^2)
=\sqg\na_\m ( P^{,\mn}P_{,\n}-P^{,\m}\na^2 P )\ ,&\nn
& K_{2b}\equiv\sqg(P^{,\mn}P_{,\mn}+\half (\na^2P)^2)\ ,\ 
g_\mn\frac{\del}{\del g_\mn (x)}\intfx K_{2b}=+2 J_{2b}\ ,&\nn
& J_{2c}\equiv\sqg\na_\m\na_\n(P^{,\m}P^{,\n})\ ,\ 
K_{2c}\equiv \sqg(R^\mn P_{,\m}P_{,\n}-\frac{1}{6}RP_{,\m}P^{,\m}) ,&\nn
&g_\mn\frac{\del}{\del g_\mn (x)}\intfx K_{2c}=+ J_{2c}\ ,&\nn
&J_{2d}\equiv \sqg\na_\m(P^{,\mn}P_{,\n})\ ,\ 
K_{2d}\equiv \sqg (P_{,\mn}P^{,\mn}-\half (\na^2 P)^2)\ ,&\nn
&g_\mn\frac{\del}{\del g_\mn (x)}\intfx K_{2d}=+2 J_{2d}\ ,&\nn
& J_3\equiv\sqg (\na^2P\cdot P_{,\m}P^{,\m}+2P^{,\m}P^{,\n}P_{,\mn})
=\sqg\na^\n ( P_{,\n}P_{,\m}P^{,\m})\ ,&\nn
& K_3\equiv\sqg P^{,\mn}P_{,\m}P_{,\n}\ ,\ 
g_\mn\frac{\del}{\del g_\mn (x)}\intfx K_3= \half J_3\ ,&\nn
& J_1\equiv\sqg(3\na^4P-\na^\la(RP_{,\la}))\ ,\ 
K_{1}\equiv\sqg R\na^2P\ ,\ 
g_\mn\frac{\del}{\del g_\mn (x)}\intfx K_{1}= J_1\ .& 
%
%
\eea
$J$'s  are trivial terms, while $K$'s  are
the corresponding counterterms. 
We note all trivial terms are total derivative ones.
But total derivative terms are not always trivial ones.
$I_{1}$ is such an example
\footnote{
$I_1$ looks like the Euler term in the sense that its Weyl transformation
is similar, as pointed out below (\ref{inv2}), and it is the surface term
which depends only on the topology.
}
.

\vspace{2cm}
{\large Appendix E. Conformal Anomaly for 2D Dilaton Coupled Scalar }

\vs{0.5}
For completeness we write below the conformal anomaly for 2D 
dilaton coupled scalar with the action:
\bea
\label{E.1}
S=\half\int d^2x\sqg\e^{-2\p}(\na\vp)^2\pr
\eea
The notation is the same as in Sec.2 of the text.
The calculation of the corresponding trace anomaly has been
done actually in Ref.\cite{ENO} (see later works\cite{BH,NO,SI97,KLV})
with the following result
\bea
\label{E.2}
T=\frac{\sqg}{24\pi}\{ -R-6(\na \p)^2+6\na^2\p \}\pr
\eea
The structure of the above anomaly consists of  the Euler term, 
$\sqg R$, a trivial term, $\sqg\na^2\p$, and a conformally invariant
term, $\sqg(\na\p)^2$. So the structure is the same as 4D 
dilaton coupled scalar in Sec.2 of the text.

The analysis given in above works leads to  
different values for the coefficient of the trivial term. In particulary,
the result of ref.\cite{BH} does not coincide with the calculations 
of refs.\cite{ENO,NO,SI97,KLV}.
The discrepancy occurs because of the difference in the choice of
the ultraviolet regularization, the system operator
and the measure\cite{SI97}. The most crucial point is, 
as far as we are studying the anomaly problem in the
perturbative quantization of the matter fields only,
we have no control over the purely gravity-dilaton sector.
It means the corresponding coefficient appears to be ambigious.
We have the freedom to introduce a finite counterterm
$\sqg R\p$ in the local gravity-dilaton action. 
(As for the the most general form of such a local action,
see Ref.\cite{OS91}.) This changes the coefficient of the trivial
term freely due to  relation:
\bea
\label{E.3}
g_\mn\frac{\del}{\del g_\mn (x)}\int d^2x\sqg R\p=\sqg \na^2\p\pr
\eea
Hence, the coefficient of the trivial term is ambigious (actually, it is 
defined by the normalization condition).
 Looking at the situation in the 2D 
black hole physics, the radiation amplitude is determined
by the conformal anomaly equation 
and conservation of the energy-
momentum tensor.
The trivial term equally contributes to its determination.
If we regard the radiation amplitude to be physical, it appears that
we have to treat the the matter-gravity-dilaton system both
nonperturbatively and on an equal footing with all fields.
\vs 1
\begin{flushleft}
{\bf Acknowledgements}
\end{flushleft}
The authors thank K.Fujikawa and R.Kantowski for reading the manuscript and
some remarks. 
SI thanks R.Endo for some comments. SDO is grateful to I.Buchbinder 
and S.Nojiri for related discussions. 

\vs 1


\end{document}